\documentclass[twocolumn,prb,unsortedaddress,floatfix,amssymb]{revtex4}

\usepackage{times}
\usepackage{graphicx}

\begin{document}

\title{Coarse-Grained Model for Phospholipid\,/\,Cholesterol Bilayer}

\author{Teemu Murtola}
\author{Emma Falck}
\affiliation{
  Laboratory of Physics and Helsinki Institute of Physics,
  Helsinki University of Technology,
  P.\,O. Box 1100, FIN--02015 HUT, Finland}

\author{Michael Patra}
\author{Mikko Karttunen}
\affiliation{
  Biophysics and Statistical Mechanics Group,
  Laboratory of Computational Engineering,
  Helsinki University of Technology,
  P.\,O. Box 9203, FIN--02015 HUT, Finland}

\author{Ilpo Vattulainen}
\affiliation{
  Laboratory of Physics~and Helsinki Institute of Physics,
  Helsinki University of Technology,
  P.\,O. Box 1100, FIN--02015 HUT, Finland}

\date{June 30, 2004}

\begin{abstract}
We construct a coarse-grained (CG) model for 
dipalmitoylphosphatidylcholine (DPPC)\,/\,cholesterol bilayers 
and apply it to large-scale simulation studies of lipid membranes. 
Our CG model is a two-dimensional representation of the membrane, 
where the individual lipid and sterol molecules are described 
by point-like particles. The effective intermolecular interactions 
used in the model are systematically derived from detailed 
atomic-scale molecular dynamics simulations using the Inverse 
Monte Carlo technique, which guarantees that the radial distribution 
properties of the CG model are consistent with those given by the 
corresponding atomistic system. We find that the coarse-grained 
model for the DPPC\,/\,cholesterol bilayer is substantially more 
efficient than atomistic models, providing a speed-up of 
approximately eight orders of magnitude. The results are in favor of
formation of cholesterol-rich and cholesterol-poor domains at 
intermediate cholesterol concentrations, in agreement with the 
experimental phase diagram of the system. We also explore the 
limits of the novel coarse-grained model, and discuss the general 
validity and applicability of the present approach.
\end{abstract}

\maketitle

\section{Introduction}
Cell membranes are remarkably flexible and durable 
structures enclosing and protecting the contents of cells
or organelles.\cite{Gennis89,Bloom91,Sac95,Merz96,%
alberts.bray.ea:molecular,Katsaras01} 
They are, however, by no means mere casings, but bustling 
hubs, where signaling, recognition, and transport take place. 
The fact that a huge variety of cellular processes are 
governed by membranes makes them a fascinating and 
biologically relevant example of soft and distinctly thin 
interfaces.

The complexity and biological relevance of membranes is 
largely due to the variety of proteins and lipids that are 
their main building blocks. It is intriguing that there 
are typically more than a hundred different lipid species 
in a given type of biological membrane, all assumed to 
have some particular purpose. \cite{Gennis89,Sac95} 
Instead of being static, membranes are highly dynamic 
and characterized by distinct phases and internal fluctuating 
structures, thus allowing proteins to function under 
non-equilibrium conditions. \cite{Bloom91} Understanding 
the overall properties of membranes is therefore a major 
challenge involving studies over large scales in time and 
space: starting from the atomistic and molecular regimes 
where small-scale processes such as ion flow through channel 
proteins takes place all the way to mesoscopic and macroscopic 
regimes where large-scale processes such as phase separation 
and membrane fusion are important.

The present understanding of membrane systems is largely 
based on experimental studies, where techniques such 
as fluorescence spectroscopy, nuclear magnetic resonance, 
and various scattering methods have been used. 
\cite{Gennis89,Bloom91,Sac95,Merz96,Katsaras01} At the 
same time, experiments have been complemented by 
theoretical and computational modeling.\cite{Bloom91,Sac95,Zuc04,%
tieleman.marrink.ea:computer,feller:molecular,scott:modeling,%
Sai02a,Sai02b,Vat04a-review} 
Thanks to the interplay between the two fields, 
a more detailed understanding of membranes and their 
biological relevance is emerging.

As for computational modeling of membrane systems in the 
{\it atomistic regime}, classical molecular 
dynamics (MD) is the method of choice. 
\cite{tieleman.marrink.ea:computer,feller:molecular,scott:modeling,%
Sai02a,Sai02b,Vat04a-review} It provides detailed 
information on the structure and dynamics of individual 
lipid molecules, as well as insight into processes such 
as the complexation and hydrogen bonding properties of 
different lipid species or the effect of enzymes on membranes. 
The main limitation of the atomistic approach is the 
computational load. With currently available computer power, 
the standard size for systems is 128 lipid molecules, 
corresponding to a linear system size of about 5\,--\,7\,nm
in the bilayer plane. The duration of such a simulation is 
typically limited to about 100\,ns. A few more ambitious MD 
simulations on systems with a larger number of molecules have 
been reported, 
\cite{chiu.vasudevan.ea:structure,hofsass.lindahl.ea:molecular,Vries04} 
but the sizes are still rather modest: the largest 
MD studies of lipid bilayers contain of the order of 10$^3$ 
lipid molecules. The time scales reached in such simulations 
are currently only tens of nanoseconds.

The above limitations are problematic since many interesting 
phenomena in lipid membrane systems occur at much 
longer time and length scales. Examples of such phenomena are domain 
formation, bilayer fusion, and cooperative motions associated 
with phase changes. Domain formation is a particularly 
interesting issue, since there is plenty of experimental 
evidence pointing to the formation of lateral domains in 
many-component bilayers. \cite{Bloom91,Sac95} 
The most topical issue are {\it lipid rafts}, 
\cite{Simons97,Mayor04,Edidin03,almeida.fedorov.ea:sphingomyelin} 
which are thought to be dynamic, ordered lateral domains 
comprised mainly of phosphatidylcholine, cholesterol, 
and sphingomyelin molecules. Rafts have been suggested to be 
involved in a wide range of cellular processes including 
membrane trafficking and sorting of proteins, 
\cite{Simons97,Mayor04,Edidin03,almeida.fedorov.ea:sphingomyelin} 
which emphasizes the need to understand large-scale properties 
of membrane domains. The dimensions of these dynamic domains 
are believed to range from tens to hundreds of nanometers, 
still beyond the limits of atomistic simulations. 
To reach the necessary length scales, i.\,e., hundreds of 
nanometers, we therefore need to resort to {\it coarse-grained 
models} that employ effective interaction potentials for 
simplified molecular descriptions. 
\cite{Bloom91,mouritsen.dammann.ea:computer,Nielaba02,SoftSimu}  
To gain insight into large-scale properties of lipid membrane 
systems, the main objective is hence to develop and employ 
coarse-grained membrane models incorporating only the essential 
properties of the underlying system.

Previous studies in this field have been few, although 
recent progress is very promising. As 
for {\it semi-atomistic models} of bilayers, a number 
of interesting approaches have been suggested. 
\cite{marrink.vries.ea:coarse,Goe98,Goe99,Imp03,Gro01,Kra03a,Kra03b,%
shelley.shelley.ea:coarse,She01b,Nie03} The guiding principle 
is that small groups of atoms are represented as single 
interaction sites, thus reducing the computational 
complexity of the model. In the model by Marrink \emph{et al.} 
\cite{marrink.vries.ea:coarse} there are Lennard--Jones, 
harmonic, and electrostatic interactions between the 
coarse-grained particles, and the interaction parameters  
have been tuned to match experimental quantities such as 
heats of vaporization or densities. The systems may
be simulated using classical MD. The approach by Lipowsky 
\emph{et al.} is somewhat more phenomenological but 
similar in nature. \cite{Goe98,Goe99,Imp03}
Shelley \emph{et al.}, in turn, have employed a large 
variety of different interactions between the coarse-grained 
particles, \cite{shelley.shelley.ea:coarse,She01b,Nie03}  
and the interaction parameters have been adjusted to match 
results from both experiments and atomic simulations. 
Groot and Rabone have employed the dissipative 
particle dynamics (DPD) technique and chosen soft potentials 
between the coarse-grained particles. \cite{Gro01} The 
repulsive interaction parameters have been derived from 
compressibility and solubilities. The DPD studies by 
Groot and Rabone have been complemented by the simulations 
of Smit \emph{et al.}\cite{Kra03a,Kra03b} 
Ayton \emph{et al.} have also employed the DPD technique. 
They have used material properties from atomistic simulations 
to parameterize meso-scale and macro-scale models
of lipid bilayers and unilamellar vesicles.\cite{Ayton01,%
Ayton02a,Ayton02b,Ayton02c}

Alternatively, one can design {\it phenomenological models} 
where the number of degrees of freedom is as small as 
possible. One of the best examples of this approach is the 
work by Mouritsen \emph{et al.}\cite{miao.nielsen.ea:from,
nielsen.miao.ea:off-lattice,Nie96,Nie00,Zuc04,%
polson.vattulainen.ea:simulation} They have developed and 
used an off-lattice model where lipid and sterol molecules 
are described as hard-core particles with internal (spin-type) 
degrees of freedom. This approach has allowed them to 
design models whose phase diagrams are in qualitative 
agreement with experimental ones for phosphatidylcholine 
(PC)\,/\,cholesterol and PC\,/\,lanosterol systems. 
\cite{miao.nielsen.ea:from,nielsen.miao.ea:off-lattice,Nie96,Nie00,Zuc04}
Additionally, the models have been successful in describing 
lateral diffusion in PC\,/\,sterol bilayer mixtures. 
\cite{polson.vattulainen.ea:simulation} 
The work by Mouritsen \emph{et al.} demonstrates that purely 
phenomenological models can be very useful in accessing scales 
larger than those within reach of atomistic simulation 
techniques. On the other hand, due to their phenomenological 
nature, the scope of these models may be limited.

In general, there is no single method of constructing 
models for mesoscopic or macroscopic phenomena, and each case 
has to be considered separately. Hence, systematic and general 
approaches that simplify the construction of coarse-grained 
models and reduce the number of phenomenological and tunable 
parameters would be of great interest. A generally useful 
approach be easily extendable or modifiable to 
describe different kinds of systems.

A promising candidate is the Inverse Monte Carlo technique (IMC). 
\cite{lyubartsev.laaksonen:calculation,lyubartsev.karttunen.ea:on} 
It allows one to derive {\it all} effective interaction potentials 
systematically from atomic-level information such that the most 
relevant structural properties of the atomic-level system are 
reproduced by the coarse-grained model. This approach has been 
used in other soft matter systems. For example Lyubartsev 
\emph{et al.} \cite{lyubartsev.laaksonen:calculation,%
lyubartsev.karttunen.ea:on,Lyu96} have used IMC for constructing 
a coarse-grained model for sodium and chloride ions in water, 
and for describing the binding of different alkali ions to DNA. 
\cite{Lyu99} The approach employed by Shelley \emph{et al.}\ 
is also, in part, based on the central ideas of the IMC method. 
\cite{shelley.shelley.ea:coarse}  
Notably, IMC is not only systematic, but also highly adjustable, as
the level of coarse-graining and the number of degrees 
of freedom can be tuned. It is thus possible to 
choose how large scales are to be studied as well as
how much detail is to be included.

\begin{figure}[t]
\includegraphics[scale=0.8]{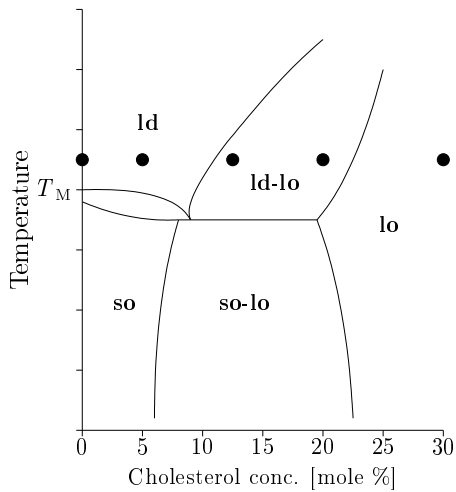}
\caption{\label{figure:phasediag}
Sketch of experimental phase diagram  for 
the DPPC\,/\,cholesterol system.\cite{vist.davis:phase} 
At high temperatures and low cholesterol concentrations, 
there is a liquid-disordered (\textbf{ld}) phase, which is 
a fluid-phase characterized by lipid acyl chains with a 
high degree of conformational disorder. When the temperature 
is lowered, the system goes through the main phase transition 
at $T_\mathrm{M}  \approx 311$\,K to a solid-ordered 
(\textbf{so}) phase. The \textbf{so} phase is 
essentially a solid phase in which acyl chains are 
conformationally ordered and the positions of the molecules 
are characterized by translational order in the bilayer plane. 
Finally, at high cholesterol concentrations, there is 
a liquid-ordered (\textbf{lo}) phase, characterized both 
by a high degree of acyl chain ordering and the lack of 
translational order found in the \textbf{ld} phase. 
At intermediate cholesterol 
concentrations there are wide \textbf{ld}-\textbf{lo} and 
\textbf{so}-\textbf{lo} coexistence regions.
The dots represent the concentrations at which the atomic-scale 
molecular dynamics simulations have been performed. An additional 
MD simulation was performed at 50\,\% cholesterol.}
\end{figure}

In the present study, we apply the IMC approach\cite{%
lyubartsev.laaksonen:calculation,lyubartsev.karttunen.ea:on} 
to construct a coarse-grained (CG) 
model for a lipid bilayer containing dipalmitoylphosphatidylcholine 
(DPPC) and cholesterol. This system was chosen because 
DPPC is one of the most studied phospholipids, and cholesterol 
is the most important sterol molecule found in plasma 
membranes of eukaryotic cells. Further, the system has a 
rich and interesting phase behavior \cite{vist.davis:phase} 
characterized by three main phases (see Fig.~\ref{figure:phasediag}).
It has been suggested that at 
certain temperatures and cholesterol concentrations the system 
might display cholesterol-rich domains \cite{slotte:lateral} 
or superlattice domains. \cite{cannon.heath.ea:time-resolved} 

We first performed extensive atomic-level MD simulations
of the DPPC\,/\,cholesterol system at six different cholesterol
concentrations.\cite{falck.patra.ea:lessons} 
The results of these simulations agree well with  
experiments and other simulations. Based on these atomistic 
considerations, we construct a coarse-grained model in which we 
describe each molecule by a single point-like particle moving 
in a two-dimensional plane. We use the IMC technique to derive 
effective interaction potentials for the coarse-grained particles.
These interactions are constructed such that the CG model 
reproduces the radial distribution functions (RDFs) calculated 
from the atomic-level MD simulations. Because the RDFs can be 
used to characterize the phase behavior, the model should also 
qualitatively reproduce the phase behavior of the microscopic 
model.

Using the coarse-grained model, we study 
DPPC\,/\,cholesterol bilayers with cholesterol concentrations 
varying from 0\,\% to 50\,\%. We first validate the CG model by 
comparing its behavior to that of the atomic-scale model. As 
the degree of coarse-graining is very high, the 
model allows us to study the properties of the bilayer on 
length scales of the order of 100\,nm along the plane of the 
membrane. The computational gain can be approximated to be around
eight orders of magnitude compared to the atomistic MD case.

We find that the coarse-graining approach can provide plenty of insight 
into large-scale properties of many-component membrane systems. 
In this case it allows us to observe 
formation of cholesterol-rich and cholesterol-poor domains at 
intermediate cholesterol concentrations, in agreement with the 
experimental phase diagram of the system. \cite{vist.davis:phase} 
We also explore the limitations of the model, 
and discuss its general validity as well as possible future applications.

\section{Molecular Dynamics Simulation Details
         \label{section:mddetails}}

The underlying MD simulations have been described in detail 
elsewhere, \cite{falck.patra.ea:lessons} and only a brief summary 
is given here. We simulated fully hydrated lipid bilayer 
systems consisting of 128 macromolecules, i.\,e., DPPCs and 
cholesterols, and 3655 water molecules. The simulations were 
performed at six cholesterol molar fractions: 0\,\%, 4.7\,\%, 
12.5\,\%, 20.3\,\%, 29.7\,\%, and 50\,\%.
These concentrations are indicated in the phase diagram of 
Fig.~\ref{figure:phasediag}. The duration of each simulation 
was 100\,ns and the linear sizes of the systems in the plane 
of the bilayer were between 5 and 7\,nm.

The starting point for the simulations was a united atom model 
for a fully hydrated pure DPPC bilayer that has been validated 
previously. \cite{patra.karttunen.ea:molecular,%
patra.karttunen.ea:lipid,tieleman.berendsen:molecular} 
The parameters for bonded and non-bonded interactions for DPPC 
molecules were taken from a study of a pure DPPC bilayer, 
\cite{berger.edholm.ea:molecular} and partial charges from the 
underlying model description. \cite{tieleman.berendsen:molecular} 
The cholesterol force field was taken from an earlier study. 
\cite{holtje.forster.ea:molecular}

As an initial configuration for the pure DPPC bilayer we used 
the final structure of run E discussed in Ref.~\onlinecite{%
tieleman.berendsen:molecular}. For systems containing cholesterol, 
the initial configurations were constructed by replacing randomly 
selected DPPC molecules with cholesterols. The same number of 
DPPC molecules was replaced in each of the two monolayers. To 
fill the small voids left by replacing DPPC molecules by somewhat 
smaller cholesterol molecules, the system was equilibrated in 
several stages. \cite{falck.patra.ea:lessons}

The MD simulations were performed at a temperature 
$T = 323$\,K using the GROMACS molecular simulation package. 
\cite{lindahl.hess.ea:gromacs}  The 
main phase transition temperature for a pure DPPC bilayer 
is $T_\mathrm{M} \approx 311$\,K,\cite{Vis90} indicating that the MD 
simulations have been conducted above $T_\mathrm{M}$. 
The time step for the simulations was chosen to be 2.0\,fs. 
Long-range electrostatic interactions were handled using the 
Particle-Mesh Ewald method. \cite{essman.perera.ea:smooth} 
After the initial equilibration we performed 100\,ns of MD 
in the $NpT$ ensemble with a Berendsen thermostat and barostat 
\cite{berendsen.postma.ea:molecular} for each cholesterol 
concentration. For all systems up to and including the 
cholesterol concentration of 29.7\,\%, a simulation lasting 
100\,ns guarantees good sampling of the phase space. The results 
for 50\,\% cholesterol should be regarded with some caution as 
the diffusion of the DPPC and cholesterol molecules is already 
quite slow. \cite{falck.patra.ea:lessons}

\section{Coarse-Grained Model
         \label{section:cgmodel}}

Using the MD simulations as a basis, we have constructed 
a coarse-grained model for a DPPC\,/\,cholesterol bilayer. 
Since the main goal of the present project is to study 
the {\it large-scale structural properties of the bilayer}, 
the degree of coarse-graining must be high. A way 
to achieve this goal is to describe each DPPC and cholesterol 
molecule by its center-of-mass (CM) position. The 
macromolecules are taken to be single point-like particles 
that move and interact in two dimensions with continuous 
coordinates. At the same time, the solvent degrees of freedom 
have been integrated out altogether, i.\,e., the model contains 
no explicit water.

Let us briefly list the assumptions we have made in 
constructing the CG model. To start with, we consider 
a lipid bilayer as a purely two-dimensional sheet comprised of 
two weakly interacting leaflets. For this reason, we focus 
on one monolayer only. This assumption is well justified 
since interdigitation in DPPC\,/\,cholesterol bilayers is 
minor. \cite{falck.patra.ea:lessons}  Consequently, 
the friction between the two leaflets is weak and they 
can be regarded as largely independent from each other. 
Furthermore, we neglect the out-of-plane fluctuations of the 
bilayer and assume it to be strictly planar. Such fluctuations 
decrease when the cholesterol concentration is increased, 
\cite{hofsass.lindahl.ea:molecular} making this a reasonable 
assumption especially at higher cholesterol concentrations.

Due to its coarse grained nature, our model is dissipative. 
This stems mainly from the fact that the water molecules 
are not included in the CG model. Further, in constructing 
the model the conformational degrees of freedom of 
the macromolecules have been integrated out. Yet another 
reason is that we consider one of the leaflets rather than 
the whole membrane. If we were to study dynamical phenomena, 
the dynamics should be chosen such that there are both stochastic 
and dissipative force components describing those degrees of 
freedom that have been excluded from the CG model. As we will 
focus on structural quantities of the membrane system, we will 
not have to worry about the choice of dynamics. Instead, we 
use the Metropolis Monte Carlo (MC) technique. \cite{Frenkel-Smit} 
The question of implementing realistic dynamics to the 
model is considered in more detail at the 
end of Sect.~\ref{section:discussion}.

As for the interactions between the point-like DPPC and 
cholesterol particles, we assume they can be adequately 
described using pairwise, radially symmetric effective 
potentials. The effective interactions are computed as 
follows. From the atomistic MD simulations we calculate radial 
distribution functions for the CM positions of the molecules. 
To link our coarse-grained model to the atomic-level system, 
we require that the coarse-grained model accurately reproduces 
these RDFs. This is accomplished by constructing the effective 
interaction potentials using the Inverse Monte Carlo method. 
\cite{lyubartsev.laaksonen:calculation,lyubartsev.karttunen.ea:on} 
In principle, also other canonical averages than the RDFs 
could be used as an input. However, the RDFs calculated from 
the atomic-scale MD simulations are easy to compute, and more 
importantly, give a detailed structural description of the 
system in the plane of the bilayer at short length scales. 
Because the RDFs can be used for characterizing the phase 
behavior of the system, the coarse-grained model should 
at least qualitatively reproduce the phase behavior of the 
original atomic-scale system.

In addition to the above, we fix the area per molecule in the 
CG model to be the same as the average area per molecule 
calculated from the atomistic MD simulations. The MC simulations 
will therefore be conducted in the canonical ensemble.

\section{Model Construction and Validation
         \label{section:construction_and_validation}}

\subsection{Radial Distribution Functions and 
  Areas per Molecule from Atomistic MD Simulations}

\begin{figure}[t]
\includegraphics[scale = 1]{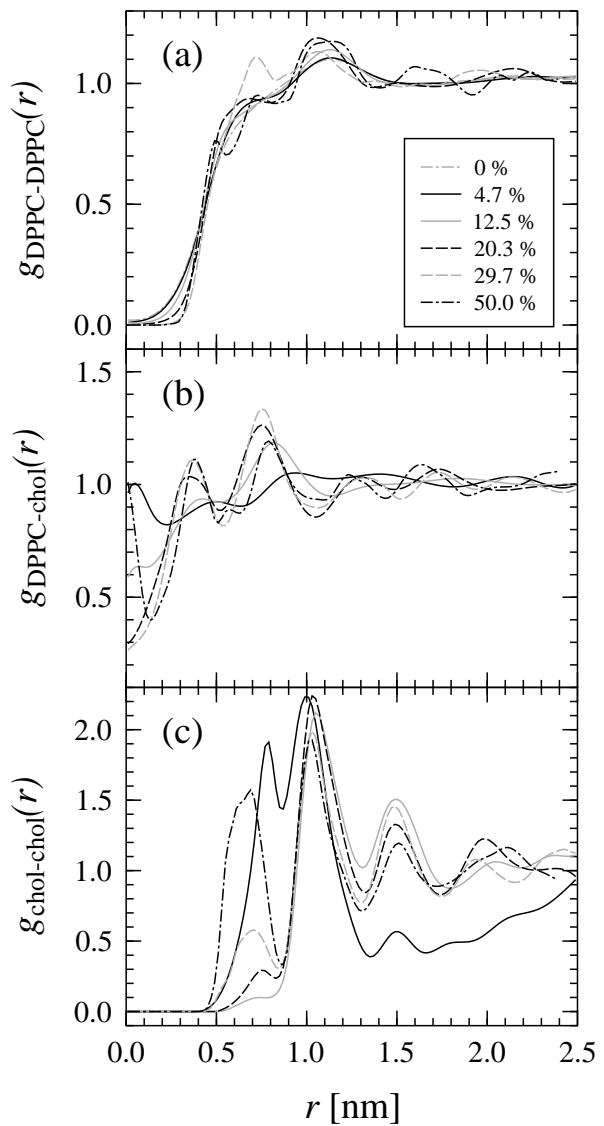}
\caption{\label{figure:mdrdfs}
Radial distribution functions calculated from MD
simulations for (a)~DPPC-DPPC, (b)~DPPC-cholesterol and 
(c) cholesterol-cholesterol pairs. The RDFs are calculated 
from the center-of-mass positions of the molecules, which 
have been projected to the plane of the bilayer.}
\end{figure}

For the construction of the coarse-grained model, we need to 
obtain the radial distribution functions for the center-of-mass 
positions of the molecules. Figure~\ref{figure:mdrdfs} shows 
the RDFs calculated from the atomic-scale MD simulations. 
Before calculating the RDFs, the CM positions have been 
projected to the plane of the bilayer. The two monolayers 
have been considered separately and the resulting RDFs are 
averages of the two. The first 20\,ns of the MD data have 
been discarded to allow the system to equilibrate fully. 
\cite{falck.patra.ea:lessons}  After 20\,ns the area per
molecule has converged for all cholesterol concentrations, 
\cite{falck.patra.ea:lessons} and the radial distribution 
functions show no systematic changes. The radial distribution 
functions were calculated up to 2.5\,nm for concentrations 
lower than 50\,\%. At the highest concentration of 50\,\%, 
the linear size of the system in the bilayer plane is 
occasionally below 5\,nm, and 
therefore in this case the RDFs were cut off at 2.4\,nm. 
The errors of the RDFs can be estimated to be of the order 
of a few percent, with somewhat higher errors at low cholesterol 
concentrations for the RDFs involving cholesterol. To minimize 
the effect of random errors on the Inverse Monte Carlo procedure, 
we used a spline-fitting procedure designed for noisy data 
\cite{thijsse.hollanders.ea:practical} to smooth the RDFs.

For all concentrations the RDFs indicate liquid-like behavior. At 
short length scales there are broad peaks and at 
larger $r$ the functions approach unity. In other words, although 
there is short-range order, there are no signs of 
of long-range order characteristic 
to solid-like phases. This confirms that we are probing the 
region of the phase diagram where the system is in the 
\textbf{ld}, \textbf{lo}, or coexisting \textbf{ld} and 
\textbf{lo} phases (see the points marked in 
Fig.~\ref{figure:phasediag}). As the cholesterol concentration 
is increased, the radial distribution functions change more 
or less systematically. The peaks in the DPPC-DPPC distribution 
become sharper, manifesting an increase in the lateral 
short-range ordering.  Further, more peaks appear at larger 
$r$, which means that the range of 
the ordering increases slightly. Similar effects are observed 
in the case of the cholesterol-cholesterol RDFs, although 
the 4.7\,\% concentration deviates somewhat from the general 
trend. For the DPPC-cholesterol distribution the changes are 
not quite as systematic. Nonetheless, when the cholesterol 
concentration grows, the range of ordering seems to 
increase slightly and the peaks generally become sharper.

A notable feature of the RDFs, especially for the 
DPPC-cholesterol pairs, is the fact that some RDFs do not 
approach zero at the origin. This is because the RDFs have 
been calculated for the CM positions of the molecules, which 
have been projected to the plane of the bilayer. It is not 
too difficult to imagine a situation where the projected CM 
positions of a rigid, short cholesterol molecule and a DPPC 
molecule with long, flexible tails are essentially on top 
of each other.

\begin{figure}[t]
\includegraphics[scale = 1]{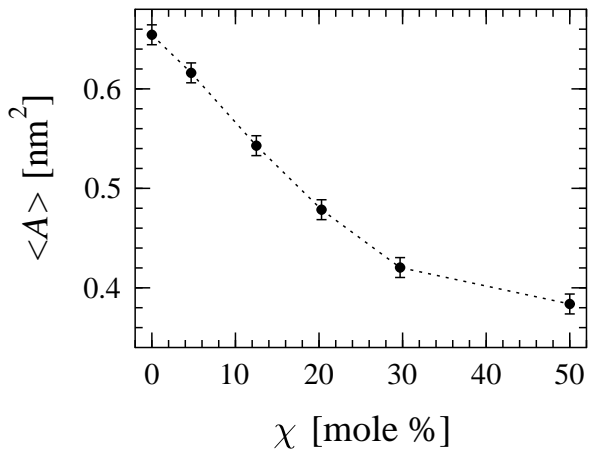}
\caption{\label{figure:areapermolecule}
Average area per molecule as a function of cholesterol 
concentration, calculated from MD simulations.\cite{falck.patra.ea:lessons}}
\end{figure}

In addition to the RDFs, we need the average area per molecule from 
the MD simulations to fix the area per molecule in the 
coarse-grained model. The average areas per molecule at 
different cholesterol concentrations have been published 
previously, \cite{falck.patra.ea:lessons} and are shown in 
Fig.~\ref{figure:areapermolecule} for reference. The area 
per molecule for a given configuration has been computed by 
dividing the size of the simulation box in the bilayer plane 
by the number of molecules in one monolayer. The area per 
molecule decreases monotonically with an increasing 
cholesterol concentration, in agreement with other 
simulations \cite{chiu.jakobsson.ea:cholesterol-induced,%
hofsass.lindahl.ea:molecular} and related experiments.\cite{McC03}

\subsection{Constructing Effective Potentials}

Based on the RDFs, we have 
constructed effective interaction 
potentials for our coarse-grained particles using the 
IMC method. For each cholesterol concentration, the RDFs
calculated from the MD simulations were given as an input 
to the IMC procedure to obtain the effective interactions.

Due to finite size effects, in some cases 
the RDFs calculated from the MD simulations deviate from unity 
at the cutoff. This results in discontinuities in the effective 
potentials at the cutoff. To handle these, we applied a simple 
shifting scheme from 2.0\,nm to the cutoff distance to adjust 
the potentials such that they approach zero smoothly at 
the cutoff. The approach used is essentially similar to that 
presented in Ref.~\onlinecite{gromacs-manual}. The main 
difference is that in the present case shifting is applied 
to the potential rather than to the force.

The IMC does not give reasonable estimates for 
the effective potentials at short interparticle distances 
where the RDFs vanish. In these regions, we replaced the 
potential given by the IMC 
method by polynomials such that the potential and its 
first derivative are continuous at the edge of the region. 
Finally, the effective potentials were smoothed using the 
same spline-fitting procedure \cite{thijsse.hollanders.ea:practical} 
as was used for the RDFs to reduce statistical noise. We have 
verified that the potentials are not sensitive to the details 
of the process of obtaining them. Thus the above changes can 
be made without seriously altering the resulting RDFs.

\begin{figure}[t]
\includegraphics[scale = 1]{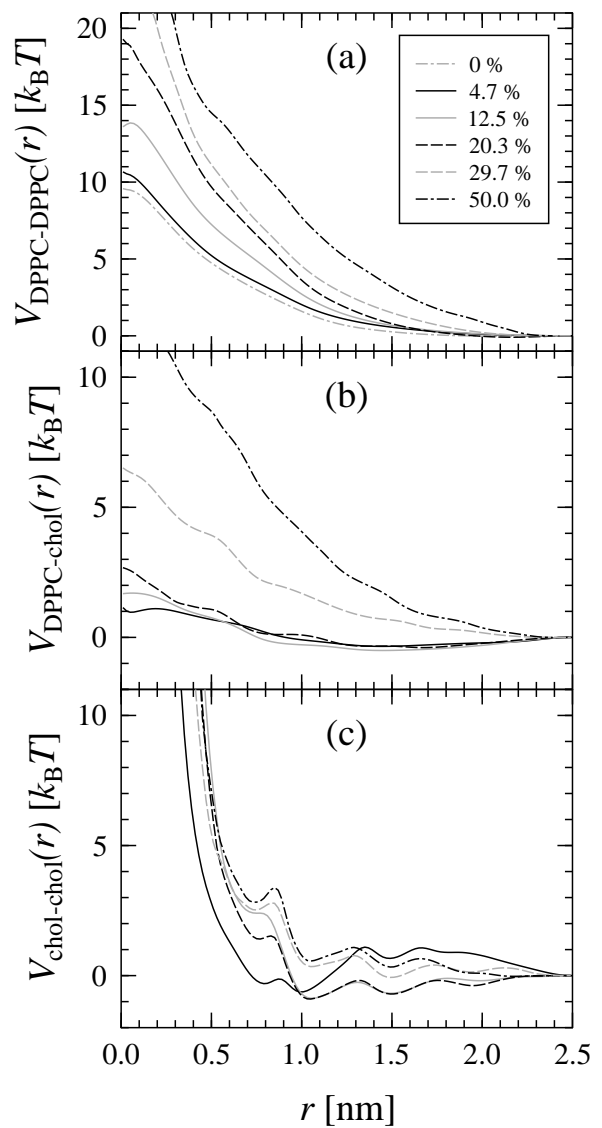}
\caption{\label{figure:potentials}
Effective potentials for different pairs of coarse-grained
particles: 
(a)~DPPC-DPPC, 
(b)~DPPC-cholesterol, and
(c)~cholesterol-cholesterol.}
\end{figure}

Figure\,\ref{figure:potentials} shows the computed effective 
interaction potentials. Due to the high level of 
coarse-graining, the potentials are soft. The DPPC-DPPC 
and DPPC-cholesterol potentials 
become systematically more repulsive with an 
increasing cholesterol concentration, and the very small 
attractive component present at low concentrations is lost 
at higher concentrations. For the cholesterol-cholesterol 
potentials, the behavior is more complex. For 29.7~\% and 
50.0~\% concentrations, the interaction is mostly repulsive, 
but for 12.5~\% and 20.3~\% there is weak attraction for 
$r \gtrsim 0.9$~nm up to the cutoff. For 4.7~\%, the interaction 
is again repulsive for $r \gtrsim 1.1$~nm, but now there is 
a weak attraction for small separations. In addition, the 
cholesterol-cholesterol potentials have multiple minima, 
whereas the other potentials are much simpler.

\subsection{Validation}

As any model, the coarse-grained model should be validated. 
This can be done by comparing the results it generates 
to the results from the MD simulations. By construction, 
the CG model should reproduce the short-range 
structural properties of the atomic-scale model.

\begin{figure}[t]
\includegraphics[scale = 1]{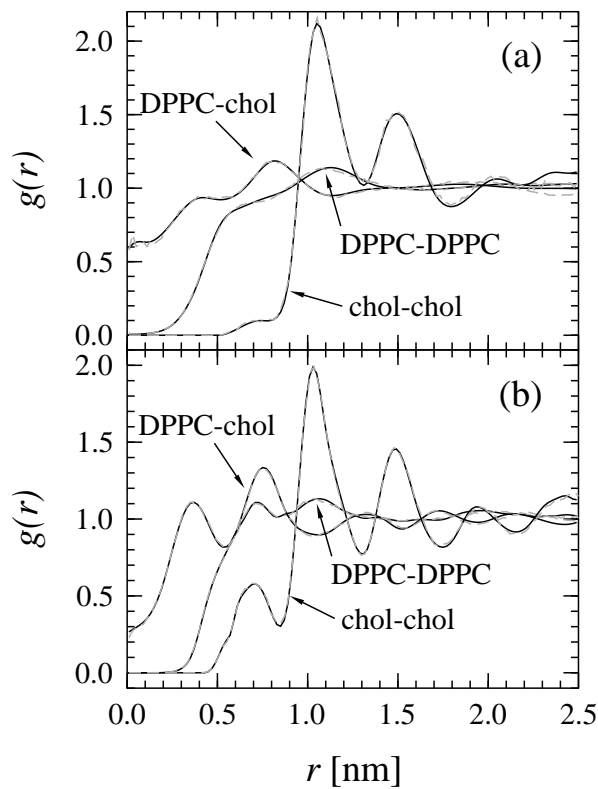}
\caption{\label{figure:cgrdfs}
Comparison between radial distribution functions 
calculated from MD simulations (solid black lines) 
and from CG model (dashed grey lines). Two different 
cholesterol concentrations are shown: 
(a) 12.5\,\% and (b) 29.7\,\%. The results for 
the other concentrations are similar.}
\end{figure}

Figure~\ref{figure:cgrdfs} shows a comparison between 
RDFs calculated from the MD simulations and those 
obtained from the CG model using canonical MC. The CG 
simulations contained the same number of particles as 
the MD simulations. We show only a few, selected cholesterol 
concentrations, but for other concentrations the results are 
similar: the agreement between the results from MD 
and CG is excellent at all concentrations, as it should. 
The minor differences near the cutoff arise from the use 
of a shifting function for the potentials. Without the 
shift, the lines coincide up to the cutoff, but in some 
cases there is a small discontinuity in the RDFs at the cutoff.

\begin{figure}[t]
\includegraphics[scale = 1]{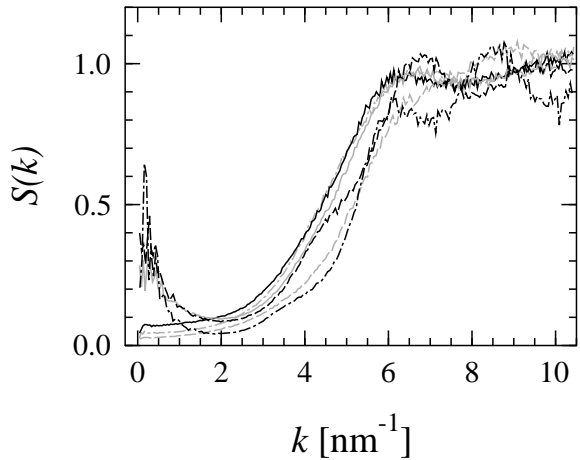}
\caption{\label{figure:stastr_summary}
Total circularly averaged static structure factors 
computed from CG simulations at different cholesterol 
concentrations. The curves are labeled as 
in Fig.~\ref{figure:mdrdfs}.}
\end{figure}

We have also calculated the static structure factors computed 
over all pairs of particles on a two-dimensional grid. To this 
end, consider a set of particles $i=1,\ldots,N$ whose positions 
are $\vec{r}_i$. Then the static structure factor defined as 
\begin{equation} 
S(\vec{k}\,) = \frac{1}{N^2}
             \langle \, 
             \sum_{i=1}^N \sum_{j=1}^N 
             \exp [ -i \, \vec{k} \cdot (\vec{r}_{j} - \vec{r}_{i}) ]  
             \, \rangle 
\end{equation} 
is given in terms of the reciprocal vector $\vec{k}$. 
The $S(\vec{k}\,)$  have been calculated for a system whose 
linear size varies between 120\,--\,160\,nm. This is 24 times 
the size of the original system studied by MD. In all cases the structure 
factors were found to be radially symmetric. The circularly 
averaged structure factors, $S(k)$,  for different cholesterol 
concentrations are shown in Fig.~\ref{figure:stastr_summary}. 
These curves are discussed in detail in 
Sect.~\ref{section:largescale}.  At this point it is 
sufficient to note that these calculations confirm that the 
system is isotropic at all concentrations and that there is 
no long-range solid-like order. The system is in 
a fluid-like state as it should.

\section{Behavior at Large Length Scales
         \label{section:largescale}}

When the system size is increased from that of the MD 
simulations, several new phenomena are observed. The 
simulations described below were mostly performed on 
systems containing 36\,864 particles, corresponding to 
linear sizes 24 times those of the original MD simulations. 
Hence, the linear sizes of the systems were 120\,--\,160\,nm, 
depending on the concentration. A typical simulation 
required 50\,--\,100\,CPU hours on a desktop computer.

\begin{figure}[t]
\includegraphics[scale = 1]{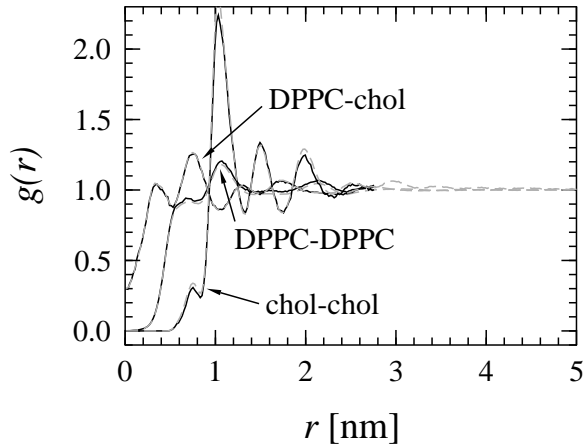}
\caption{\label{figure:cgrdf_size}
Changes in radial distribution functions when the size
of the simulation box is increased at 20.3\,\% cholesterol. 
Black solid lines correspond to a small system with 64 particles 
and dashed grey lines to a system with a linear size 24 
times larger.}
\end{figure}

When the system size is increased, there are some minor 
changes in the radial distribution functions. These are 
illustrated in Fig.~\ref{figure:cgrdf_size} for the 20.3\,\% 
cholesterol concentration. For other concentrations the
 results are similar. These changes are most probably 
a finite-size effect caused by the small sizes of the 
original atomic-scale systems. It is very likely that 
if the MD simulations could be performed on larger systems, 
similar changes in the RDFs should take place. The figure 
also shows that the RDFs rapidly approach unity at large distances.

\begin{figure*}[t]
\includegraphics[scale = 1]{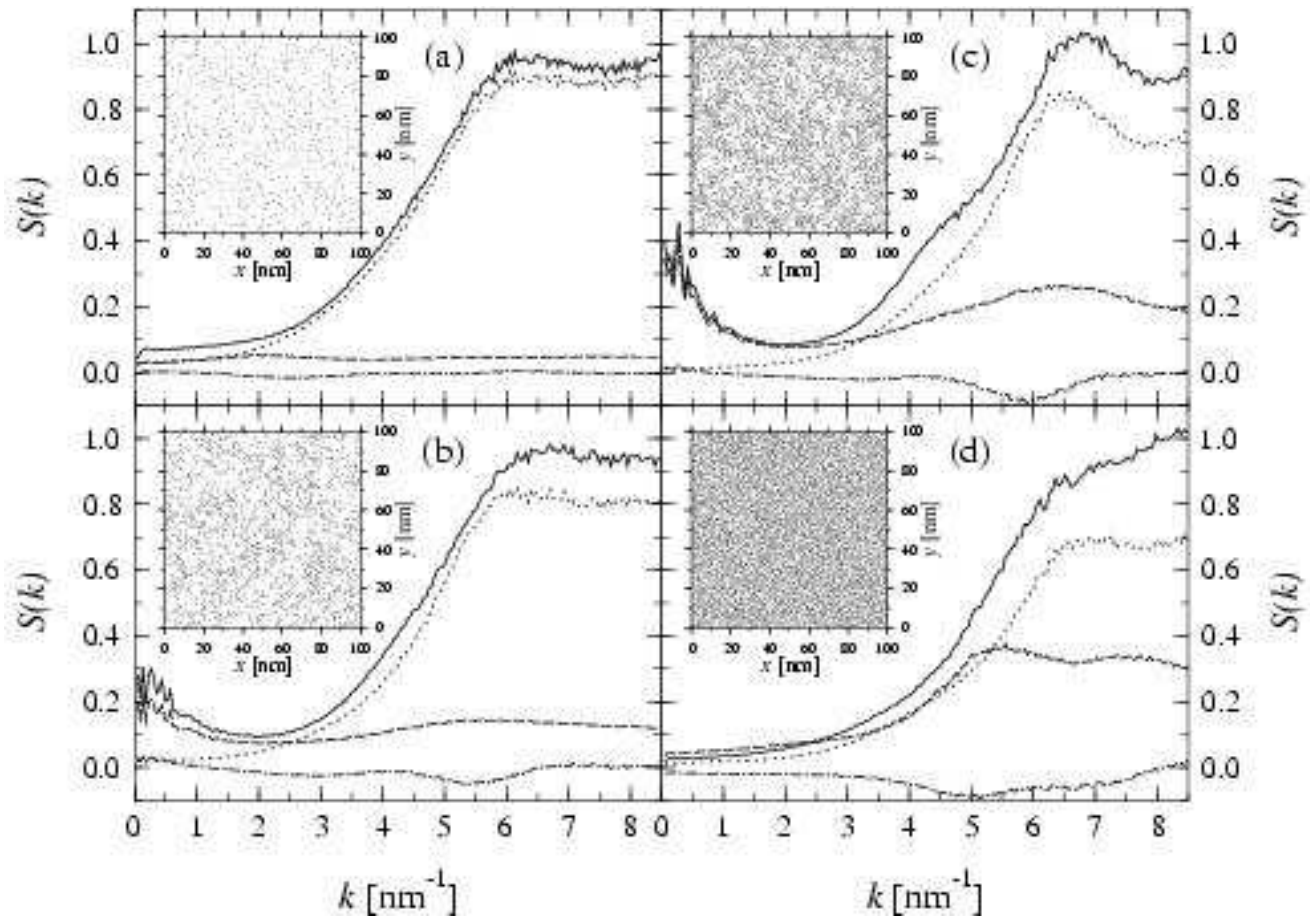}
\caption{\label{figure:stastr}
Static structure factors calculated for different
cholesterol concentrations: 
(a)~4.7\,\%, 
(b)~12.5\,\%, 
(c)~20.3\,\%, and 
(d)~29.7\,\%. 
In addition to the total structure factor computed over 
all pairs of particles (solid line), structure factors 
calculated  over DPPC-DPPC (dotted), cholesterol-cholesterol 
(dashed) and DPPC-cholesterol (dash-dotted) pairs are shown. 
Also a snapshot of the system is shown for each concentration. 
In the snapshots, each cholesterol molecule is represented 
as a single dot, while DPPC molecules are not shown.}
\end{figure*}

To study possible large-scale organization, we have looked 
at the static structure factors calculated at different 
cholesterol concentrations. They 
are shown in Fig.~\ref{figure:stastr} together with snapshots 
of the system at each concentration. The structure factors 
have been calculated over all pairs of molecules, and also 
separately for DPPC-DPPC, cholesterol-cholesterol, and 
DPPC-cholesterol pairs.  To ensure that the static structure 
factors do not depend on the initial configuration,
we have rerun the calculations using several different 
initial states. We find that different initial configurations 
lead to results that are consistent with each other.

At 12.5\,\% and 20.3\,\% cholesterol, the snapshots in 
Fig.~\ref{figure:stastr} suggest that there are domains 
where the local concentration of cholesterol is higher 
than in other regions. At these concentrations the 
cholesterol-cholesterol structure factor shows a rather 
wide peak at small $k$, supporting our interpretation of the 
presence of cholesterol-rich and cholesterol-poor domains. 
The maximum of the peak corresponds to length scales 
of the order of 20\,nm or more. A more 
precise analysis is difficult, since even for the largest 
systems we have studied, with linear sizes of approximately 
280\,nm, the peak is rather broad, and the small $k$ side 
of the peak is not fully clear due to fluctuations at 
large length scales.

We have, however, studied how the
peak behaves if the system size is varied by calculating 
the static structure factors for four 
systems with linear sizes 6, 12, 24 and 48 times that of 
the original MD simulations. In this finite size scaling analysis, we 
found no clear change in either the position or shape of the peak.

The formation of cholesterol-rich and cholesterol-poor 
domains at intermediate cholesterol concentrations is in agreement 
with the phase diagram, see Fig.~\ref{figure:phasediag}. 
According to the phase diagram we should, at $T = 323$\,K, expect 
both the MD and the CG model to display coexisting \textbf{ld} 
and \textbf{lo} phases. In the present case, we cannot directly 
distinguish between the \textbf{ld} and \textbf{lo} phases, 
since we have not included the ordering of the DPPC tails in 
the model. However, we might argue as follows to establish
that the cholesterol-rich phases should be \textbf{lo}, 
while the cholesterol-poor must be \textbf{ld}. The 
ordering effect of cholesterol on the phospholipid tails has 
been clearly demonstrated: the higher the cholesterol 
concentration, the more ordered the tails. 
\cite{falck.patra.ea:lessons,hofsass.lindahl.ea:molecular} 
It is thus plausible that the tails should be more ordered 
in the cholesterol-rich regions.

To study domain formation in more detail, we have computed 
probability distribution functions for finding square domains 
with a linear size $\ell$. Several different system sizes 
$\ell$ were considered. If there were domains of clearly 
different compositions, the distribution should display two 
peaks. These studies do not provide direct support for 
formation of domains (data not shown). At each concentration, 
and for each $\ell$,
the distribution is very close to a Gaussian. However, with 
12.5\,\% and 20.3\,\% cholesterol the variance of the 
distribution is clearly larger than for 4.7\,\% 
or 29.7\,\% cholesterol. Thus, it is possible that the observed 
domains are caused by large-scale fluctuations in the cholesterol 
concentration. Furthermore, at 12.5\,\% the distribution is not 
quite symmetric for $\ell \lesssim 20$\,nm. This may be due 
to two peaks that are located very close to each other.

We have also investigated the local concentrations of 
different types of molecules in the vicinity of a given 
type of molecule as in a study by de~Vries
\emph{et~al.} \cite{vries.mark.ea:binary}  More specifically, 
we have studied the respective numbers of different types 
of molecules among the $n$ nearest neighbors of DPPC and 
cholesterol molecules, for different values of $n$. The 
average number of molecules of type A among the $n$ nearest 
neighbors of a molecule of type B was compared to the value 
in a purely random configuration. In all cases the average 
number of cholesterol molecules in the vicinity of another 
cholesterol molecule was somewhat smaller than it should 
be in a random configuration. For $n=6$, there were no 
significant differences between different cholesterol 
concentrations. For larger $n$, say $n=15$ or $n=30$, the 
difference between the value from an actual simulation 
and the value from a random configuration was clearly 
smaller at the concentrations where we observe cholesterol-rich 
and cholesterol-poor domains than at other concentrations. 
This observation may be interpreted as support to the 
existence of domains. The reason why the difference is not 
visible in the case of $n=6$ is probably that such a small 
neighborhood represents the area from which other cholesterol 
molecules are largely excluded by the repulsive interactions 
between the cholesterol molecules.

Based mainly on fluorescence measurements, it has been 
proposed that at certain cholesterol concentrations the 
cholesterol molecules could adopt a regular arrangement 
in parts of the bilayer.\cite{somerharju.virtanen.ea:lateral} 
We have studied whether or not our coarse-grained model 
shows such organization for a few of the proposed ``magical'' 
concentrations. The nearest-neighbor analysis described 
above suggests that finding a cholesterol molecule in the 
immediate vicinity of another cholesterol molecule is lower 
compared to a random configuration, whereas the probability 
of finding a cholesterol molecule next to a DPPC molecule 
is higher. Such a situation would occur if there were 
superlattices in the system. Nevertheless, we have seen no 
evidence for any regular lattice-like ordering of cholesterols. 
Despite this conclusion, we are looking 
forward to further studies using other models. In our model 
the DPPC particles are radially symmetric, while in reality 
the two hydrocarbon tails of the molecules may be important 
for the occurrence of superlattices, if they do exist.

In the case of 50\,\% cholesterol the situation is more 
complex. Visual inspection of snapshots clearly indicates 
that the DPPC and cholesterol molecules phase separate. 
All other studies we have performed support this view. There 
are at least two alternative explanations for this phenomenon. 
First, it is possible that the atomistic MD simulations do not, at this 
high cholesterol concentration characterized by a very small 
lateral diffusion coefficient, adequately sample the phase 
space. If this is the case, it may lead to errors
in the RDFs extracted from the atomic-level simulations. As 
a consequence, the effective potentials given by the IMC method 
would not describe the true behavior. On the other hand, 
there is experimental evidence for the formation of crystalline 
cholesterol domains at very high cholesterol concentrations. 
\cite{bach.wachtel:phospholipid} 
Also this could be the reason for the observed phase 
separation. Based on the current simulations, it is difficult 
to say which of these, if indeed any, is the case.

\section{Discussion
         \label{section:discussion}}

The coarse-graining approach described here has several 
advantages. It is a systematic method to generate
a mesoscale model that is linked to atomic-scale information. 
It is adjustable, as it allows the user to control the 
level of coarse-graining and the number of degrees of freedom 
to be included in the model. The approach is general in 
the sense that it can be applied to a wide range of systems. 
The only prerequisite is that we should somehow acquire radial 
distribution functions for pairs of the coarse-grained particles. 
These are available from atomic-scale simulations, but in 
certain cases also experiments could be used to derive the RDFs.

The most important advantage, however, is speed. Monte Carlo 
simulations using the coarse-grained model are several orders 
of magnitude faster than MD simulations of the original 
atomic-scale model. The speedup can be estimated 
by considering the decay time of the fluctuations 
in the area of a single molecule, where the area is calculated, 
through Voronoi analysis.\cite{falck.patra.ea:lessons,
patra.karttunen.ea:lipid} For the atomic-scale model, this 
decay time is approximately 0.7\,ns, 
\cite{patra.karttunen.ea:lipid} whereas in the case of 
the CG model similar decay is observed after one MC step. 
The CPU time required for generating a 0.7\,ns 
trajectory for the atomic-scale system using MD is around 70\,h, 
whereas for a CG model of the same linear size as the atomic-scale 
system,  approximately 25\,000 Monte Carlo steps can be taken 
during a minute. Thus the speedup can be estimated to be eight 
orders of magnitude.

Despite all advantages, there are limitations. An important point 
is that any problems in the atomic-scale molecular 
dynamics simulations are transferred 
to the coarse-grained model. In particular, if there is any 
artificial ordering in the atomic-scale simulations, either 
due to a small system size, poor sampling of the phase space, 
or sloppy treatment of electrostatic interactions, 
\cite{patra.karttunen.ea:molecular,patra.karttunen.ea:lipid} 
the radial distribution functions will be erroneous. As 
a consequence, the effective interaction potentials will be affected.

Another limitation may be the 
concentration-dependence of the effective potentials, see 
Fig.~\ref{figure:potentials}. It is not trivial in what
concentration ranges the effective potentials are valid. 
The worst case scenario is that when the cholesterol
concentration is altered ever so slightly, the interaction 
potentials must be rederived starting from time-consuming 
atomic-scale molecular dynamics simulations. We have
investigated whether the effective potentials 
can be used at concentrations other than those at which they 
were determined. The results suggest that they may be used 
at nearby concentrations, but if phase boundaries are
crossed, problems will arise.  For instance, when using the 
effective potentials determined for the system with 29.7\,\% 
cholesterol, a system in the \textbf{lo} phase, to 
simulate a system at 20.3\,\% cholesterol, which should be in 
the coexistence region, no long-range structure appears.  When,
on the other hand, potentials determined at 20.3\,\% cholesterol 
are used with 29.7\,\% cholesterol, the long-range structure 
is still present at the higher concentration, although the 
peak is more shallow. The situation is similar when we compare 
the 4.7\,\% and 12.5\,\% concentrations. When the potentials 
determined at 12.5\,\% are used for 20.3\,\% cholesterol or 
vice versa, i.\,e., both concentrations are in the coexistence 
region, we still observe domains. The detailed form of the 
static structure factor will, however, be altered. 
We may therefore conclude that the effective potentials 
cannot be used for mapping the precise phase 
boundaries of a given system. Similar conclusions can be 
drawn from a study of the temperature dependence of the 
effective interaction potentials.

An additional problem is the implementation of dynamics.
In this study, we considered only structural quantities, 
and needed not to worry about the choice of dynamics. 
To study lateral diffusion it is necessary
to incorporate realistic dynamics into the system. It is 
well-known that this can be very difficult in the case 
of coarse-grained models. In addition to the MC studies 
reported here, we attempted to study lateral diffusion 
and include dynamics into our model. In choosing the
dynamics, the following criteria should be met.
First, as explained in Sect.~\ref{section:cgmodel}, the 
dissipative nature of the system should be taken into 
account. In addition, the simulations should be run in 
the canonical ensemble. We tested several thermostats that 
comply to the above requirements: standard and generalized 
Brownian dynamics,\cite{ermak.buckholtz:numerical} and an 
approach similar to Andersen's thermostat for temperature 
coupling. \cite{andersen:molecular} None of these could be 
tuned to give realistic dynamics for the system with 
rattling-in-the-cage movement and separate jump events. 
Both the standard Brownian dynamics approach and the 
Andersen scheme require very large friction or coupling
parameters to give diffusion coefficients of the same 
order of magnitude as those obtained from the molecular 
dynamics simulations, or alternatively, to match the decay 
times of velocity autocorrelation functions. Such high 
values of the parameters completely determine the dynamics 
at short time scales, and the interactions between the 
particles only give rise to small corrections at long 
time scales. The generalized Brownian dynamics can be used 
to force the short-time dynamics to match that of molecular 
dynamics simulations, but this does not mend the problem of 
unrealistic dynamics at longer time scales. We thus found 
no simple way of implementing realistic dynamics into the 
coarse-grained model. Reducing the degree of coarse-graining 
and introducing more detail into the model, e.\,g., by 
including the tails of DPPC and cholesterol molecules, might 
help to solve the problem. The presence of tails would 
greatly increase the friction between the molecules, and 
possibly allow for entanglements. Both effects would help 
in slowing down the unrealistically fast dynamics.

The results motivate further studies of the 
coarse-graining approach. There are several possible directions 
in which the model could be developed. Several of these 
could be realized without additional MD simulations. A possible 
line of development is the inclusion of the conformational 
degrees of freedom of the DPPC molecules in the model. This 
could be done in the spirit of the model by Nielsen \emph{et al.}, 
\cite{nielsen.miao.ea:off-lattice} i.\,e., by giving each DPPC 
molecule two possible states: an ordered and a disordered state. 
We would have three kinds of particles and a total of six 
pairwise potentials to determine. Another possible modification 
would be to include the two tails of the lipid molecules as 
separate particles and possibly model the head group as a third 
particle. Results from simulations of such models could be 
compared to the present study to assess the possible benefits 
of such modifications, and to gain further insight into the 
coarse-graining process. Additionally, to better understand 
the underlying reasons for domain formation, it is natural
to ask what the relative roles of entropic and energetic 
contributions in this process are. Since the close-packed 
areas of the CG particles are not well defined, a study of this 
broad and non-trivial issue is beyond the scope of this work 
and will be discussed elsewhere.

The coarse-graining approach presented here could also be 
applied to other lipid bilayer systems to compare their behavior 
to the DPPC\,/\,cholesterol bilayer. A particularly interesting 
system is the sphingomyelin (SM)\,/\,cholesterol bilayer, 
and ultimately the PC\,/\,SM\,/\,cholesterol ternary mixture. 
The study of these systems at large length scales would be 
particularly interesting, because experimental results 
suggest the existence of specific interactions between SM 
and cholesterol molecules. \cite{almeida.fedorov.ea:sphingomyelin}
These interactions are believed to forward the formation of 
domains and lipid rafts at high concentrations of SM and 
cholesterol. This line of development is limited by the 
computational cost of the underlying MD simulations. 
Currently this confines us to study quite simple bilayer 
systems. With increasing computer power and improvements in 
existing simulation methods, studies of ternary mixtures of 
lipids, as well as studies of systems containing membrane 
proteins, are becoming feasible.

\section{Summary and conclusions
         \label{section:summary}}

We have used a systematic approach for 
constructing coarse-grained models for DPPC\,/\,cholesterol 
bilayers. The central ingredient is the application of the 
Inverse Monte Carlo method, which can be used for finding 
effective interactions such that the coarse-grained model 
reproduces given radial distribution functions. The approach 
allows easy tuning of the level of coarse-graining, and it can 
be applied to a wide range of systems.

The radial distribution functions given as input to the 
IMC method have been extracted from detailed atomic-level 
molecular dynamics simulations. The effective interactions 
found using the IMC method can then be used to simulate the 
system on much longer length scales. We have found that the 
coarse-grained model thus constructed is in favor of the 
formation of cholesterol-rich and cholesterol-poor domains 
at intermediate cholesterol concentrations, in 
agreement with the phase diagram of the system. We have also 
explored the limitations of the constructed coarse-grained model.

As for further studies, it would be interesting to see 
how modifications such as the inclusion of the 
conformational degrees of freedom of the DPPC tails would 
influence the behavior of the model. Similar models 
could also be constructed for other many-component lipid bilayer 
systems, and their behavior compared to the present study. 
This could yield valuable new information on both the systems 
under study and the suitability of the IMC method for 
coarse-graining in general.

\vspace*{.1cm}

\begin{acknowledgments}

This work has, in part, been supported by the Academy of Finland 
through its Center of Excellence Program (T.\,M, E.\,F., I.\,V.), 
the National Graduate School in Materials Physics (E.\,F.), 
the Academy of Finland Grant Nos. 80246 (I.\,V.), 54113 and 00119 (M.\,K.), 
and 125495 (T.\,M.), and by the European Union 
through the Marie Curie fellowship HPMF--CT--2002--01794 (M.\,P.). 
We would also like to thank the Finnish IT Center for Science 
and the HorseShoe (DCSC) supercluster computing facility at the 
University of Southern Denmark for computer resources. Finally, 
we wish to thank Peter Lindqvist for his help with the Voronoi 
analysis.
\end{acknowledgments}


\begin{thebibliography}{69}
\expandafter\ifx\csname natexlab\endcsname\relax\def\natexlab#1{#1}\fi
\expandafter\ifx\csname bibnamefont\endcsname\relax
  \def\bibnamefont#1{#1}\fi
\expandafter\ifx\csname bibfnamefont\endcsname\relax
  \def\bibfnamefont#1{#1}\fi
\expandafter\ifx\csname citenamefont\endcsname\relax
  \def\citenamefont#1{#1}\fi
\expandafter\ifx\csname url\endcsname\relax
  \def\url#1{\texttt{#1}}\fi
\expandafter\ifx\csname urlprefix\endcsname\relax\def\urlprefix{URL }\fi
\providecommand{\bibinfo}[2]{#2}
\providecommand{\eprint}[2][]{\url{#2}}

\bibitem[{\citenamefont{Gennis}(1989)}]{Gennis89}
\bibinfo{author}{\bibfnamefont{R.~B.} \bibnamefont{Gennis}},
  \emph{\bibinfo{title}{Biomembranes: Molecular Structure and Function}}
  (\bibinfo{publisher}{Springer-Verlag}, \bibinfo{address}{New York},
  \bibinfo{year}{1989}).

\bibitem[{\citenamefont{Bloom et~al.}(1991)\citenamefont{Bloom, Evans, and
  Mouritsen}}]{Bloom91}
\bibinfo{author}{\bibfnamefont{M.}~\bibnamefont{Bloom}},
  \bibinfo{author}{\bibfnamefont{E.}~\bibnamefont{Evans}}, \bibnamefont{and}
  \bibinfo{author}{\bibfnamefont{O.~G.} \bibnamefont{Mouritsen}},
  \bibinfo{journal}{Q. Rev. Biophys.} \textbf{\bibinfo{volume}{24}},
  \bibinfo{pages}{293} (\bibinfo{year}{1991}).

\bibitem[{\citenamefont{Lipowsky and Sackmann}(1995)}]{Sac95}
\bibinfo{editor}{\bibfnamefont{R.}~\bibnamefont{Lipowsky}} \bibnamefont{and}
  \bibinfo{editor}{\bibfnamefont{E.}~\bibnamefont{Sackmann}}, eds.,
  \emph{\bibinfo{title}{Structure and Dynamics of Membranes: From Cells to
  Vesicles}} (\bibinfo{publisher}{Elsevier}, \bibinfo{address}{Amsterdam},
  \bibinfo{year}{1995}).

\bibitem[{\citenamefont{K.~M.~Merz and Roux}(1996)}]{Merz96}
\bibinfo{editor}{\bibfnamefont{J.}~\bibnamefont{K.~M.~Merz}} \bibnamefont{and}
  \bibinfo{editor}{\bibfnamefont{B.}~\bibnamefont{Roux}}, eds.,
  \emph{\bibinfo{title}{Biological Membranes: A Molecular Perspective from
  Computation and Experiment}} (\bibinfo{publisher}{Birh\"auser},
  \bibinfo{address}{Boston}, \bibinfo{year}{1996}).

\bibitem[{\citenamefont{Alberts et~al.}(1994)\citenamefont{Alberts, Bray,
  Lewis, Raff, Roberts, and Watson}}]{alberts.bray.ea:molecular}
\bibinfo{author}{\bibfnamefont{B.}~\bibnamefont{Alberts}},
  \bibinfo{author}{\bibfnamefont{D.}~\bibnamefont{Bray}},
  \bibinfo{author}{\bibfnamefont{J.}~\bibnamefont{Lewis}},
  \bibinfo{author}{\bibfnamefont{M.}~\bibnamefont{Raff}},
  \bibinfo{author}{\bibfnamefont{K.}~\bibnamefont{Roberts}}, \bibnamefont{and}
  \bibinfo{author}{\bibfnamefont{J.~D.} \bibnamefont{Watson}},
  \emph{\bibinfo{title}{Molecular Biology of the Cell}}
  (\bibinfo{publisher}{Garland Publishing}, \bibinfo{address}{New York},
  \bibinfo{year}{1994}), \bibinfo{edition}{3rd} ed.

\bibitem[{\citenamefont{Katsaras and Gutberlet}(2001)}]{Katsaras01}
\bibinfo{editor}{\bibfnamefont{J.}~\bibnamefont{Katsaras}} \bibnamefont{and}
  \bibinfo{editor}{\bibfnamefont{T.}~\bibnamefont{Gutberlet}}, eds.,
  \emph{\bibinfo{title}{Lipid Bilayers: Structure and Interactions}}
  (\bibinfo{publisher}{Springer-Verlag}, \bibinfo{address}{Berlin},
  \bibinfo{year}{2001}).

\bibitem[{\citenamefont{Zuckermann et~al.}(2004)\citenamefont{Zuckermann,
  Bloom, Ipsen, Miao, Mouritsen, Nielsen, Polson, Thewalt, Vattulainen, and
  Zhu}}]{Zuc04}
\bibinfo{author}{\bibfnamefont{M.~J.} \bibnamefont{Zuckermann}},
  \bibinfo{author}{\bibfnamefont{M.}~\bibnamefont{Bloom}},
  \bibinfo{author}{\bibfnamefont{J.~H.} \bibnamefont{Ipsen}},
  \bibinfo{author}{\bibfnamefont{L.}~\bibnamefont{Miao}},
  \bibinfo{author}{\bibfnamefont{O.~G.} \bibnamefont{Mouritsen}},
  \bibinfo{author}{\bibfnamefont{M.}~\bibnamefont{Nielsen}},
  \bibinfo{author}{\bibfnamefont{J.}~\bibnamefont{Polson}},
  \bibinfo{author}{\bibfnamefont{J.}~\bibnamefont{Thewalt}},
  \bibinfo{author}{\bibfnamefont{I.}~\bibnamefont{Vattulainen}},
  \bibnamefont{and} \bibinfo{author}{\bibfnamefont{H.}~\bibnamefont{Zhu}},
  \bibinfo{journal}{Methods Enzym.} \textbf{\bibinfo{volume}{383}},
  \bibinfo{pages}{198} (\bibinfo{year}{2004}).

\bibitem[{\citenamefont{Tieleman et~al.}(1997)\citenamefont{Tieleman, Marrink,
  and Berendsen}}]{tieleman.marrink.ea:computer}
\bibinfo{author}{\bibfnamefont{D.~P.} \bibnamefont{Tieleman}},
  \bibinfo{author}{\bibfnamefont{S.~J.} \bibnamefont{Marrink}},
  \bibnamefont{and} \bibinfo{author}{\bibfnamefont{H.~J.~C.}
  \bibnamefont{Berendsen}}, \bibinfo{journal}{Biochim. Biophys. Acta}
  \textbf{\bibinfo{volume}{1331}}, \bibinfo{pages}{235} (\bibinfo{year}{1997}).

\bibitem[{\citenamefont{Feller}(2000)}]{feller:molecular}
\bibinfo{author}{\bibfnamefont{S.~E.} \bibnamefont{Feller}},
  \bibinfo{journal}{Curr. Opin. Coll. Interface Sci.}
  \textbf{\bibinfo{volume}{5}}, \bibinfo{pages}{217} (\bibinfo{year}{2000}).

\bibitem[{\citenamefont{Scott}(2002)}]{scott:modeling}
\bibinfo{author}{\bibfnamefont{H.~L.} \bibnamefont{Scott}},
  \bibinfo{journal}{Curr. Opin. Struct. Biol.} \textbf{\bibinfo{volume}{12}},
  \bibinfo{pages}{495} (\bibinfo{year}{2002}).

\bibitem[{\citenamefont{Saiz et~al.}(2002)\citenamefont{Saiz, Bandyopadhyay,
  and Klein}}]{Sai02a}
\bibinfo{author}{\bibfnamefont{L.}~\bibnamefont{Saiz}},
  \bibinfo{author}{\bibfnamefont{S.}~\bibnamefont{Bandyopadhyay}},
  \bibnamefont{and} \bibinfo{author}{\bibfnamefont{M.~L.} \bibnamefont{Klein}},
  \bibinfo{journal}{Biosci. Rep.} \textbf{\bibinfo{volume}{22}},
  \bibinfo{pages}{151} (\bibinfo{year}{2002}).

\bibitem[{\citenamefont{Saiz and Klein}(2002)}]{Sai02b}
\bibinfo{author}{\bibfnamefont{L.}~\bibnamefont{Saiz}} \bibnamefont{and}
  \bibinfo{author}{\bibfnamefont{M.~L.} \bibnamefont{Klein}},
  \bibinfo{journal}{Acc. Chem. Res.} \textbf{\bibinfo{volume}{35}},
  \bibinfo{pages}{482} (\bibinfo{year}{2002}).

\bibitem[{\citenamefont{Vattulainen and Karttunen}(2005, in
  press)}]{Vat04a-review}
\bibinfo{author}{\bibfnamefont{I.}~\bibnamefont{Vattulainen}} \bibnamefont{and}
  \bibinfo{author}{\bibfnamefont{M.}~\bibnamefont{Karttunen}}, in
  \emph{\bibinfo{booktitle}{Computational Nanotechnology}}, edited by
  \bibinfo{editor}{\bibfnamefont{M.}~\bibnamefont{Rieth}} \bibnamefont{and}
  \bibinfo{editor}{\bibfnamefont{W.}~\bibnamefont{Schommers}}
  (\bibinfo{publisher}{Americal Scientific Press}, \bibinfo{year}{2005, in
  press}).

\bibitem[{\citenamefont{Chiu et~al.}(2003)\citenamefont{Chiu, Vasudevan,
  Jakobsson, Mashl, and Scott}}]{chiu.vasudevan.ea:structure}
\bibinfo{author}{\bibfnamefont{S.~W.} \bibnamefont{Chiu}},
  \bibinfo{author}{\bibfnamefont{S.}~\bibnamefont{Vasudevan}},
  \bibinfo{author}{\bibfnamefont{E.}~\bibnamefont{Jakobsson}},
  \bibinfo{author}{\bibfnamefont{R.~J.} \bibnamefont{Mashl}}, \bibnamefont{and}
  \bibinfo{author}{\bibfnamefont{H.~L.} \bibnamefont{Scott}},
  \bibinfo{journal}{Biophys. J.} \textbf{\bibinfo{volume}{85}},
  \bibinfo{pages}{3624} (\bibinfo{year}{2003}).

\bibitem[{\citenamefont{Hofs\"a\ss et~al.}(2003)\citenamefont{Hofs\"a\ss,
  Lindahl, and Edholm}}]{hofsass.lindahl.ea:molecular}
\bibinfo{author}{\bibfnamefont{C.}~\bibnamefont{Hofs\"a\ss}},
  \bibinfo{author}{\bibfnamefont{E.}~\bibnamefont{Lindahl}}, \bibnamefont{and}
  \bibinfo{author}{\bibfnamefont{O.}~\bibnamefont{Edholm}},
  \bibinfo{journal}{Biophys. J.} \textbf{\bibinfo{volume}{84}},
  \bibinfo{pages}{2192} (\bibinfo{year}{2003}).

\bibitem[{\citenamefont{de~Vries
  et~al.}(2004{\natexlab{a}})\citenamefont{de~Vries, Mark, and
  Marrink}}]{Vries04}
\bibinfo{author}{\bibfnamefont{A.~H.} \bibnamefont{de~Vries}},
  \bibinfo{author}{\bibfnamefont{A.~E.} \bibnamefont{Mark}}, \bibnamefont{and}
  \bibinfo{author}{\bibfnamefont{S.~J.} \bibnamefont{Marrink}},
  \bibinfo{journal}{J. Am. Chem. Soc.} \textbf{\bibinfo{volume}{126}},
  \bibinfo{pages}{4488} (\bibinfo{year}{2004}{\natexlab{a}}).

\bibitem[{\citenamefont{Simons and Ikonen}(1997)}]{Simons97}
\bibinfo{author}{\bibfnamefont{K.}~\bibnamefont{Simons}} \bibnamefont{and}
  \bibinfo{author}{\bibfnamefont{E.}~\bibnamefont{Ikonen}},
  \bibinfo{journal}{Nature} \textbf{\bibinfo{volume}{387}},
  \bibinfo{pages}{569} (\bibinfo{year}{1997}).

\bibitem[{\citenamefont{Mayor and Rao}(2004)}]{Mayor04}
\bibinfo{author}{\bibfnamefont{S.}~\bibnamefont{Mayor}} \bibnamefont{and}
  \bibinfo{author}{\bibfnamefont{M.}~\bibnamefont{Rao}},
  \bibinfo{journal}{Traffic} \textbf{\bibinfo{volume}{5}}, \bibinfo{pages}{231}
  (\bibinfo{year}{2004}).

\bibitem[{\citenamefont{Edidin}(2003)}]{Edidin03}
\bibinfo{author}{\bibfnamefont{M.}~\bibnamefont{Edidin}},
  \bibinfo{journal}{Annu. Rev. Biophys. Biomol. Struct.}
  \textbf{\bibinfo{volume}{32}}, \bibinfo{pages}{257} (\bibinfo{year}{2003}).

\bibitem[{\citenamefont{de~Almeida et~al.}(2003)\citenamefont{de~Almeida,
  Fedorov, and Prieto}}]{almeida.fedorov.ea:sphingomyelin}
\bibinfo{author}{\bibfnamefont{R.~F.~M.} \bibnamefont{de~Almeida}},
  \bibinfo{author}{\bibfnamefont{A.}~\bibnamefont{Fedorov}}, \bibnamefont{and}
  \bibinfo{author}{\bibfnamefont{M.}~\bibnamefont{Prieto}},
  \bibinfo{journal}{Biophys. J.} \textbf{\bibinfo{volume}{85}},
  \bibinfo{pages}{2406} (\bibinfo{year}{2003}).

\bibitem[{\citenamefont{Mouritsen et~al.}(1995)\citenamefont{Mouritsen,
  Dammann, Fogedby, Ipsen, Jeppesen, J{\o}rgensen, Risbo, Sabra, Sperotto, and
  Zuckermann}}]{mouritsen.dammann.ea:computer}
\bibinfo{author}{\bibfnamefont{O.~G.} \bibnamefont{Mouritsen}},
  \bibinfo{author}{\bibfnamefont{B.}~\bibnamefont{Dammann}},
  \bibinfo{author}{\bibfnamefont{H.~C.} \bibnamefont{Fogedby}},
  \bibinfo{author}{\bibfnamefont{J.~H.} \bibnamefont{Ipsen}},
  \bibinfo{author}{\bibfnamefont{C.}~\bibnamefont{Jeppesen}},
  \bibinfo{author}{\bibfnamefont{K.}~\bibnamefont{J{\o}rgensen}},
  \bibinfo{author}{\bibfnamefont{J.}~\bibnamefont{Risbo}},
  \bibinfo{author}{\bibfnamefont{M.~C.} \bibnamefont{Sabra}},
  \bibinfo{author}{\bibfnamefont{M.~M.} \bibnamefont{Sperotto}},
  \bibnamefont{and} \bibinfo{author}{\bibfnamefont{M.~J.}
  \bibnamefont{Zuckermann}}, \bibinfo{journal}{Biophys. Chem.}
  \textbf{\bibinfo{volume}{55}}, \bibinfo{pages}{55} (\bibinfo{year}{1995}).

\bibitem[{\citenamefont{Nielaba et~al.}(2002)\citenamefont{Nielaba, Mareschal,
  and Ciccotti}}]{Nielaba02}
\bibinfo{editor}{\bibfnamefont{P.}~\bibnamefont{Nielaba}},
  \bibinfo{editor}{\bibfnamefont{M.}~\bibnamefont{Mareschal}},
  \bibnamefont{and} \bibinfo{editor}{\bibfnamefont{G.}~\bibnamefont{Ciccotti}},
  eds., \emph{\bibinfo{title}{Bridging Time Scales: Molecular Simulations for
  the Next Decade}} (\bibinfo{publisher}{Springer-Verlag},
  \bibinfo{address}{Berlin}, \bibinfo{year}{2002}).

\bibitem[{\citenamefont{Karttunen et~al.}(2004)\citenamefont{Karttunen,
  Vattulainen, and Lukkarinen}}]{SoftSimu}
\bibinfo{editor}{\bibfnamefont{M.}~\bibnamefont{Karttunen}},
  \bibinfo{editor}{\bibfnamefont{I.}~\bibnamefont{Vattulainen}},
  \bibnamefont{and}
  \bibinfo{editor}{\bibfnamefont{A.}~\bibnamefont{Lukkarinen}}, eds.,
  \emph{\bibinfo{title}{Novel Methods in Soft Matter Simulations}}
  (\bibinfo{publisher}{Springer-Verlag}, \bibinfo{address}{Berlin},
  \bibinfo{year}{2004}).

\bibitem[{\citenamefont{Marrink et~al.}(2004)\citenamefont{Marrink, de~Vries,
  and Mark}}]{marrink.vries.ea:coarse}
\bibinfo{author}{\bibfnamefont{S.~J.} \bibnamefont{Marrink}},
  \bibinfo{author}{\bibfnamefont{A.~H.} \bibnamefont{de~Vries}},
  \bibnamefont{and} \bibinfo{author}{\bibfnamefont{A.~E.} \bibnamefont{Mark}},
  \bibinfo{journal}{J. Phys. Chem. B} \textbf{\bibinfo{volume}{108}},
  \bibinfo{pages}{750} (\bibinfo{year}{2004}).

\bibitem[{\citenamefont{Goetz and Lipowsky}(1998)}]{Goe98}
\bibinfo{author}{\bibfnamefont{R.}~\bibnamefont{Goetz}} \bibnamefont{and}
  \bibinfo{author}{\bibfnamefont{R.}~\bibnamefont{Lipowsky}},
  \bibinfo{journal}{J. Chem. Phys.} \textbf{\bibinfo{volume}{108}},
  \bibinfo{pages}{7397} (\bibinfo{year}{1998}).

\bibitem[{\citenamefont{Goetz et~al.}(1999)\citenamefont{Goetz, Gompper, and
  Lipowsky}}]{Goe99}
\bibinfo{author}{\bibfnamefont{R.}~\bibnamefont{Goetz}},
  \bibinfo{author}{\bibfnamefont{G.}~\bibnamefont{Gompper}}, \bibnamefont{and}
  \bibinfo{author}{\bibfnamefont{R.}~\bibnamefont{Lipowsky}},
  \bibinfo{journal}{Phys. Rev. Lett.} \textbf{\bibinfo{volume}{82}},
  \bibinfo{pages}{221} (\bibinfo{year}{1999}).

\bibitem[{\citenamefont{Imparato et~al.}(2003)\citenamefont{Imparato,
  Shillcock, and Lipowsky}}]{Imp03}
\bibinfo{author}{\bibfnamefont{A.}~\bibnamefont{Imparato}},
  \bibinfo{author}{\bibfnamefont{J.~C.} \bibnamefont{Shillcock}},
  \bibnamefont{and} \bibinfo{author}{\bibfnamefont{R.}~\bibnamefont{Lipowsky}},
  \bibinfo{journal}{Eur. Phys. J. E} \textbf{\bibinfo{volume}{11}},
  \bibinfo{pages}{21} (\bibinfo{year}{2003}).

\bibitem[{\citenamefont{Groot and Rabone}(2001)}]{Gro01}
\bibinfo{author}{\bibfnamefont{R.~D.} \bibnamefont{Groot}} \bibnamefont{and}
  \bibinfo{author}{\bibfnamefont{K.~L.} \bibnamefont{Rabone}},
  \bibinfo{journal}{Biophys. J.} \textbf{\bibinfo{volume}{81}},
  \bibinfo{pages}{725} (\bibinfo{year}{2001}).

\bibitem[{\citenamefont{Kranenburg
  et~al.}(2003{\natexlab{a}})\citenamefont{Kranenburg, Venturoli, and
  Smit}}]{Kra03a}
\bibinfo{author}{\bibfnamefont{M.}~\bibnamefont{Kranenburg}},
  \bibinfo{author}{\bibfnamefont{M.}~\bibnamefont{Venturoli}},
  \bibnamefont{and} \bibinfo{author}{\bibfnamefont{B.}~\bibnamefont{Smit}},
  \bibinfo{journal}{Phys. Rev. E} \textbf{\bibinfo{volume}{67}},
  \bibinfo{pages}{060901(R)} (\bibinfo{year}{2003}{\natexlab{a}}).

\bibitem[{\citenamefont{Kranenburg
  et~al.}(2003{\natexlab{b}})\citenamefont{Kranenburg, Venturoli, and
  Smit}}]{Kra03b}
\bibinfo{author}{\bibfnamefont{M.}~\bibnamefont{Kranenburg}},
  \bibinfo{author}{\bibfnamefont{M.}~\bibnamefont{Venturoli}},
  \bibnamefont{and} \bibinfo{author}{\bibfnamefont{B.}~\bibnamefont{Smit}},
  \bibinfo{journal}{J. Phys. Chem. B} \textbf{\bibinfo{volume}{107}},
  \bibinfo{pages}{11491} (\bibinfo{year}{2003}{\natexlab{b}}).

\bibitem[{\citenamefont{Shelley
  et~al.}(2001{\natexlab{a}})\citenamefont{Shelley, Shelley, Reeder,
  Bandyopadhyay, and Klein}}]{shelley.shelley.ea:coarse}
\bibinfo{author}{\bibfnamefont{J.~C.} \bibnamefont{Shelley}},
  \bibinfo{author}{\bibfnamefont{M.~Y.} \bibnamefont{Shelley}},
  \bibinfo{author}{\bibfnamefont{R.~C.} \bibnamefont{Reeder}},
  \bibinfo{author}{\bibfnamefont{S.}~\bibnamefont{Bandyopadhyay}},
  \bibnamefont{and} \bibinfo{author}{\bibfnamefont{M.~L.} \bibnamefont{Klein}},
  \bibinfo{journal}{J. Phys. Chem. B} \textbf{\bibinfo{volume}{105}},
  \bibinfo{pages}{4464} (\bibinfo{year}{2001}{\natexlab{a}}).

\bibitem[{\citenamefont{Shelley
  et~al.}(2001{\natexlab{b}})\citenamefont{Shelley, Shelley, Reeder,
  Bandyopadhyay, Moore, and Klein}}]{She01b}
\bibinfo{author}{\bibfnamefont{J.~C.} \bibnamefont{Shelley}},
  \bibinfo{author}{\bibfnamefont{M.~Y.} \bibnamefont{Shelley}},
  \bibinfo{author}{\bibfnamefont{R.~C.} \bibnamefont{Reeder}},
  \bibinfo{author}{\bibfnamefont{S.}~\bibnamefont{Bandyopadhyay}},
  \bibinfo{author}{\bibfnamefont{P.~B.} \bibnamefont{Moore}}, \bibnamefont{and}
  \bibinfo{author}{\bibfnamefont{M.~L.} \bibnamefont{Klein}},
  \bibinfo{journal}{J. Phys. Chem. B} \textbf{\bibinfo{volume}{105}},
  \bibinfo{pages}{9785} (\bibinfo{year}{2001}{\natexlab{b}}).

\bibitem[{\citenamefont{Nielsen et~al.}(2003)\citenamefont{Nielsen, Lopez,
  Moore, Shelley, and Klein}}]{Nie03}
\bibinfo{author}{\bibfnamefont{S.~O.} \bibnamefont{Nielsen}},
  \bibinfo{author}{\bibfnamefont{C.~F.} \bibnamefont{Lopez}},
  \bibinfo{author}{\bibfnamefont{P.~B.} \bibnamefont{Moore}},
  \bibinfo{author}{\bibfnamefont{J.~C.} \bibnamefont{Shelley}},
  \bibnamefont{and} \bibinfo{author}{\bibfnamefont{M.~L.} \bibnamefont{Klein}},
  \bibinfo{journal}{J. Phys. Chem. B} \textbf{\bibinfo{volume}{107}},
  \bibinfo{pages}{13911} (\bibinfo{year}{2003}).

\bibitem[{\citenamefont{Ayton et~al.}(2001)\citenamefont{Ayton, Bardenhagen,
  McMurtry, Sulsky, and Voth}}]{Ayton01}
\bibinfo{author}{\bibfnamefont{G.}~\bibnamefont{Ayton}},
  \bibinfo{author}{\bibfnamefont{S.~G.} \bibnamefont{Bardenhagen}},
  \bibinfo{author}{\bibfnamefont{P.}~\bibnamefont{McMurtry}},
  \bibinfo{author}{\bibfnamefont{D.}~\bibnamefont{Sulsky}}, \bibnamefont{and}
  \bibinfo{author}{\bibfnamefont{G.~A.} \bibnamefont{Voth}},
  \bibinfo{journal}{J. Chem. Phys.} \textbf{\bibinfo{volume}{114}},
  \bibinfo{pages}{6913} (\bibinfo{year}{2001}).

\bibitem[{\citenamefont{Ayton et~al.}(2002{\natexlab{a}})\citenamefont{Ayton,
  Smondyrev, Bardenhagen, McMurtry, and Voth}}]{Ayton02a}
\bibinfo{author}{\bibfnamefont{G.}~\bibnamefont{Ayton}},
  \bibinfo{author}{\bibfnamefont{A.~M.} \bibnamefont{Smondyrev}},
  \bibinfo{author}{\bibfnamefont{S.~G.} \bibnamefont{Bardenhagen}},
  \bibinfo{author}{\bibfnamefont{P.}~\bibnamefont{McMurtry}}, \bibnamefont{and}
  \bibinfo{author}{\bibfnamefont{G.~A.} \bibnamefont{Voth}},
  \bibinfo{journal}{Biophys. J.} \textbf{\bibinfo{volume}{82}},
  \bibinfo{pages}{1226} (\bibinfo{year}{2002}{\natexlab{a}}).

\bibitem[{\citenamefont{Ayton and Voth}(2002)}]{Ayton02b}
\bibinfo{author}{\bibfnamefont{G.}~\bibnamefont{Ayton}} \bibnamefont{and}
  \bibinfo{author}{\bibfnamefont{G.~A.} \bibnamefont{Voth}},
  \bibinfo{journal}{Biophys. J.} \textbf{\bibinfo{volume}{83}},
  \bibinfo{pages}{3357} (\bibinfo{year}{2002}).

\bibitem[{\citenamefont{Ayton et~al.}(2002{\natexlab{b}})\citenamefont{Ayton,
  Smondyrev, Bardenhagen, McMurtry, and Voth}}]{Ayton02c}
\bibinfo{author}{\bibfnamefont{G.}~\bibnamefont{Ayton}},
  \bibinfo{author}{\bibfnamefont{A.~M.} \bibnamefont{Smondyrev}},
  \bibinfo{author}{\bibfnamefont{S.~G.} \bibnamefont{Bardenhagen}},
  \bibinfo{author}{\bibfnamefont{P.}~\bibnamefont{McMurtry}}, \bibnamefont{and}
  \bibinfo{author}{\bibfnamefont{G.~A.} \bibnamefont{Voth}},
  \bibinfo{journal}{Biophys. J.} \textbf{\bibinfo{volume}{83}},
  \bibinfo{pages}{1026} (\bibinfo{year}{2002}{\natexlab{b}}).

\bibitem[{\citenamefont{Miao et~al.}(2002)\citenamefont{Miao, Nielsen, Thewalt,
  Ipsen, Bloom, Zuckermann, and Mouritsen}}]{miao.nielsen.ea:from}
\bibinfo{author}{\bibfnamefont{L.}~\bibnamefont{Miao}},
  \bibinfo{author}{\bibfnamefont{M.}~\bibnamefont{Nielsen}},
  \bibinfo{author}{\bibfnamefont{J.}~\bibnamefont{Thewalt}},
  \bibinfo{author}{\bibfnamefont{J.~H.} \bibnamefont{Ipsen}},
  \bibinfo{author}{\bibfnamefont{M.}~\bibnamefont{Bloom}},
  \bibinfo{author}{\bibfnamefont{M.~J.} \bibnamefont{Zuckermann}},
  \bibnamefont{and} \bibinfo{author}{\bibfnamefont{O.~G.}
  \bibnamefont{Mouritsen}}, \bibinfo{journal}{Biophys. J.}
  \textbf{\bibinfo{volume}{82}}, \bibinfo{pages}{1429} (\bibinfo{year}{2002}).

\bibitem[{\citenamefont{Nielsen et~al.}(1999)\citenamefont{Nielsen, Miao,
  Ipsen, Zuckermann, and Mouritsen}}]{nielsen.miao.ea:off-lattice}
\bibinfo{author}{\bibfnamefont{M.}~\bibnamefont{Nielsen}},
  \bibinfo{author}{\bibfnamefont{L.}~\bibnamefont{Miao}},
  \bibinfo{author}{\bibfnamefont{J.~H.} \bibnamefont{Ipsen}},
  \bibinfo{author}{\bibfnamefont{M.~J.} \bibnamefont{Zuckermann}},
  \bibnamefont{and} \bibinfo{author}{\bibfnamefont{O.~G.}
  \bibnamefont{Mouritsen}}, \bibinfo{journal}{Phys. Rev. E}
  \textbf{\bibinfo{volume}{59}}, \bibinfo{pages}{5790} (\bibinfo{year}{1999}).

\bibitem[{\citenamefont{Nielsen et~al.}(1996)\citenamefont{Nielsen, Miao,
  Ipsen, Mouritsen, and Zuckermann}}]{Nie96}
\bibinfo{author}{\bibfnamefont{M.}~\bibnamefont{Nielsen}},
  \bibinfo{author}{\bibfnamefont{L.}~\bibnamefont{Miao}},
  \bibinfo{author}{\bibfnamefont{J.~H.} \bibnamefont{Ipsen}},
  \bibinfo{author}{\bibfnamefont{O.~G.} \bibnamefont{Mouritsen}},
  \bibnamefont{and} \bibinfo{author}{\bibfnamefont{M.~J.}
  \bibnamefont{Zuckermann}}, \bibinfo{journal}{Phys. Rev. E}
  \textbf{\bibinfo{volume}{54}}, \bibinfo{pages}{6889} (\bibinfo{year}{1996}).

\bibitem[{\citenamefont{Nielsen et~al.}(2000)\citenamefont{Nielsen, Thewalt,
  Miao, Ipsen, Bloom, Zuckermann, and Mouritsen}}]{Nie00}
\bibinfo{author}{\bibfnamefont{M.}~\bibnamefont{Nielsen}},
  \bibinfo{author}{\bibfnamefont{J.}~\bibnamefont{Thewalt}},
  \bibinfo{author}{\bibfnamefont{L.}~\bibnamefont{Miao}},
  \bibinfo{author}{\bibfnamefont{J.~H.} \bibnamefont{Ipsen}},
  \bibinfo{author}{\bibfnamefont{M.}~\bibnamefont{Bloom}},
  \bibinfo{author}{\bibfnamefont{M.~J.} \bibnamefont{Zuckermann}},
  \bibnamefont{and} \bibinfo{author}{\bibfnamefont{O.~G.}
  \bibnamefont{Mouritsen}}, \bibinfo{journal}{Europhys. Lett.}
  \textbf{\bibinfo{volume}{52}}, \bibinfo{pages}{368} (\bibinfo{year}{2000}).

\bibitem[{\citenamefont{Polson et~al.}(2001)\citenamefont{Polson, Vattulainen,
  Zhu, and Zuckermann}}]{polson.vattulainen.ea:simulation}
\bibinfo{author}{\bibfnamefont{J.~M.} \bibnamefont{Polson}},
  \bibinfo{author}{\bibfnamefont{I.}~\bibnamefont{Vattulainen}},
  \bibinfo{author}{\bibfnamefont{H.}~\bibnamefont{Zhu}}, \bibnamefont{and}
  \bibinfo{author}{\bibfnamefont{M.~J.} \bibnamefont{Zuckermann}},
  \bibinfo{journal}{Eur. Phys. J. E} \textbf{\bibinfo{volume}{5}},
  \bibinfo{pages}{485} (\bibinfo{year}{2001}).

\bibitem[{\citenamefont{Lyubartsev and
  Laaksonen}(1995)}]{lyubartsev.laaksonen:calculation}
\bibinfo{author}{\bibfnamefont{A.~P.} \bibnamefont{Lyubartsev}}
  \bibnamefont{and}
  \bibinfo{author}{\bibfnamefont{A.}~\bibnamefont{Laaksonen}},
  \bibinfo{journal}{Phys. Rev. E} \textbf{\bibinfo{volume}{52}},
  \bibinfo{pages}{3730} (\bibinfo{year}{1995}).

\bibitem[{\citenamefont{Lyubartsev et~al.}(2003)\citenamefont{Lyubartsev,
  Karttunen, Vattulainen, and Laaksonen}}]{lyubartsev.karttunen.ea:on}
\bibinfo{author}{\bibfnamefont{A.~P.} \bibnamefont{Lyubartsev}},
  \bibinfo{author}{\bibfnamefont{M.}~\bibnamefont{Karttunen}},
  \bibinfo{author}{\bibfnamefont{I.}~\bibnamefont{Vattulainen}},
  \bibnamefont{and}
  \bibinfo{author}{\bibfnamefont{A.}~\bibnamefont{Laaksonen}},
  \bibinfo{journal}{Soft Materials} \textbf{\bibinfo{volume}{1}},
  \bibinfo{pages}{121} (\bibinfo{year}{2003}).

\bibitem[{\citenamefont{Lyubartsev and Laaksonen}(1996)}]{Lyu96}
\bibinfo{author}{\bibfnamefont{A.}~\bibnamefont{Lyubartsev}} \bibnamefont{and}
  \bibinfo{author}{\bibfnamefont{A.}~\bibnamefont{Laaksonen}},
  \bibinfo{journal}{J. Phys. Chem.} \textbf{\bibinfo{volume}{100}},
  \bibinfo{pages}{16410} (\bibinfo{year}{1996}).

\bibitem[{\citenamefont{Lyubartsev and Laaksonen}(1999)}]{Lyu99}
\bibinfo{author}{\bibfnamefont{A.}~\bibnamefont{Lyubartsev}} \bibnamefont{and}
  \bibinfo{author}{\bibfnamefont{A.}~\bibnamefont{Laaksonen}},
  \bibinfo{journal}{J. Chem. Phys.} \textbf{\bibinfo{volume}{111}},
  \bibinfo{pages}{11207} (\bibinfo{year}{1999}).

\bibitem[{\citenamefont{Vist and Davis}(1990{\natexlab{a}})}]{vist.davis:phase}
\bibinfo{author}{\bibfnamefont{M.~R.} \bibnamefont{Vist}} \bibnamefont{and}
  \bibinfo{author}{\bibfnamefont{J.~H.} \bibnamefont{Davis}},
  \bibinfo{journal}{Biochemistry} \textbf{\bibinfo{volume}{29}},
  \bibinfo{pages}{451} (\bibinfo{year}{1990}{\natexlab{a}}).

\bibitem[{\citenamefont{Slotte}(1995)}]{slotte:lateral}
\bibinfo{author}{\bibfnamefont{J.~P.} \bibnamefont{Slotte}},
  \bibinfo{journal}{Biochim. Biophys. Acta} \textbf{\bibinfo{volume}{1235}},
  \bibinfo{pages}{419} (\bibinfo{year}{1995}).

\bibitem[{\citenamefont{Cannon et~al.}(2003)\citenamefont{Cannon, Heath, Hyang,
  Somerharju, Virtanen, and Cheng}}]{cannon.heath.ea:time-resolved}
\bibinfo{author}{\bibfnamefont{B.}~\bibnamefont{Cannon}},
  \bibinfo{author}{\bibfnamefont{G.}~\bibnamefont{Heath}},
  \bibinfo{author}{\bibfnamefont{J.}~\bibnamefont{Hyang}},
  \bibinfo{author}{\bibfnamefont{P.}~\bibnamefont{Somerharju}},
  \bibinfo{author}{\bibfnamefont{J.~A.} \bibnamefont{Virtanen}},
  \bibnamefont{and} \bibinfo{author}{\bibfnamefont{K.~H.} \bibnamefont{Cheng}},
  \bibinfo{journal}{Biophys. J.} \textbf{\bibinfo{volume}{84}},
  \bibinfo{pages}{3777} (\bibinfo{year}{2003}).

\bibitem[{\citenamefont{Falck et~al.}()\citenamefont{Falck, Patra, Hyv\"onen,
  Karttunen, and Vattulainen}}]{falck.patra.ea:lessons}
\bibinfo{author}{\bibfnamefont{E.}~\bibnamefont{Falck}},
  \bibinfo{author}{\bibfnamefont{M.}~\bibnamefont{Patra}},
  \bibinfo{author}{\bibfnamefont{M.~T.} \bibnamefont{Hyv\"onen}},
  \bibinfo{author}{\bibfnamefont{M.}~\bibnamefont{Karttunen}},
  \bibnamefont{and}
  \bibinfo{author}{\bibfnamefont{I.}~\bibnamefont{Vattulainen}},
  \emph{\bibinfo{title}{Lessons of slicing membranes: Interplay of packing,
  free area, and lateral diffusion in phospholipid/cholesterol bilayers}},
  \bibinfo{note}{accepted to Biophys. J., preprint cond-mat/0402290 available
  at xxx.lanl.gov.}

\bibitem[{\citenamefont{Patra et~al.}(2003)\citenamefont{Patra, Karttunen,
  Hyv\"onen, Falck, Lindqvist, and Vattulainen}}]{patra.karttunen.ea:molecular}
\bibinfo{author}{\bibfnamefont{M.}~\bibnamefont{Patra}},
  \bibinfo{author}{\bibfnamefont{M.}~\bibnamefont{Karttunen}},
  \bibinfo{author}{\bibfnamefont{M.~T.} \bibnamefont{Hyv\"onen}},
  \bibinfo{author}{\bibfnamefont{E.}~\bibnamefont{Falck}},
  \bibinfo{author}{\bibfnamefont{P.}~\bibnamefont{Lindqvist}},
  \bibnamefont{and}
  \bibinfo{author}{\bibfnamefont{I.}~\bibnamefont{Vattulainen}},
  \bibinfo{journal}{Biophys. J.} \textbf{\bibinfo{volume}{84}},
  \bibinfo{pages}{3636} (\bibinfo{year}{2003}).

\bibitem[{\citenamefont{Patra et~al.}(2004)\citenamefont{Patra, Karttunen,
  Hyv\"onen, Falck, and Vattulainen}}]{patra.karttunen.ea:lipid}
\bibinfo{author}{\bibfnamefont{M.}~\bibnamefont{Patra}},
  \bibinfo{author}{\bibfnamefont{M.}~\bibnamefont{Karttunen}},
  \bibinfo{author}{\bibfnamefont{M.~T.} \bibnamefont{Hyv\"onen}},
  \bibinfo{author}{\bibfnamefont{E.}~\bibnamefont{Falck}}, \bibnamefont{and}
  \bibinfo{author}{\bibfnamefont{I.}~\bibnamefont{Vattulainen}},
  \bibinfo{journal}{J. Phys. Chem. B} \textbf{\bibinfo{volume}{108}},
  \bibinfo{pages}{4485} (\bibinfo{year}{2004}).

\bibitem[{\citenamefont{Tieleman and
  Berendsen}(1996)}]{tieleman.berendsen:molecular}
\bibinfo{author}{\bibfnamefont{D.~P.} \bibnamefont{Tieleman}} \bibnamefont{and}
  \bibinfo{author}{\bibfnamefont{H.~J.~C.} \bibnamefont{Berendsen}},
  \bibinfo{journal}{J. Chem. Phys.} \textbf{\bibinfo{volume}{105}},
  \bibinfo{pages}{4871} (\bibinfo{year}{1996}).

\bibitem[{\citenamefont{Berger et~al.}(1997)\citenamefont{Berger, Edholm, and
  Jahnig}}]{berger.edholm.ea:molecular}
\bibinfo{author}{\bibfnamefont{O.}~\bibnamefont{Berger}},
  \bibinfo{author}{\bibfnamefont{O.}~\bibnamefont{Edholm}}, \bibnamefont{and}
  \bibinfo{author}{\bibfnamefont{F.}~\bibnamefont{Jahnig}},
  \bibinfo{journal}{Biophys. J.} \textbf{\bibinfo{volume}{72}},
  \bibinfo{pages}{2002} (\bibinfo{year}{1997}).

\bibitem[{\citenamefont{H\"oltje et~al.}(2001)\citenamefont{H\"oltje,
  F\"orster, Brandt, Engels, von Rybinski, and
  H\"oltje}}]{holtje.forster.ea:molecular}
\bibinfo{author}{\bibfnamefont{M.}~\bibnamefont{H\"oltje}},
  \bibinfo{author}{\bibfnamefont{T.}~\bibnamefont{F\"orster}},
  \bibinfo{author}{\bibfnamefont{B.}~\bibnamefont{Brandt}},
  \bibinfo{author}{\bibfnamefont{T.}~\bibnamefont{Engels}},
  \bibinfo{author}{\bibfnamefont{W.}~\bibnamefont{von Rybinski}},
  \bibnamefont{and} \bibinfo{author}{\bibfnamefont{H.-D.}
  \bibnamefont{H\"oltje}}, \bibinfo{journal}{Biochim. Biophys. Acta}
  \textbf{\bibinfo{volume}{1511}}, \bibinfo{pages}{156} (\bibinfo{year}{2001}).

\bibitem[{\citenamefont{Lindahl et~al.}(2001)\citenamefont{Lindahl, Hess, and
  van~der Spoel}}]{lindahl.hess.ea:gromacs}
\bibinfo{author}{\bibfnamefont{E.}~\bibnamefont{Lindahl}},
  \bibinfo{author}{\bibfnamefont{B.}~\bibnamefont{Hess}}, \bibnamefont{and}
  \bibinfo{author}{\bibfnamefont{D.}~\bibnamefont{van~der Spoel}},
  \bibinfo{journal}{J. Mol. Mod.} \textbf{\bibinfo{volume}{7}},
  \bibinfo{pages}{306} (\bibinfo{year}{2001}).

\bibitem[{\citenamefont{Vist and Davis}(1990{\natexlab{b}})}]{Vis90}
\bibinfo{author}{\bibfnamefont{M.~R.} \bibnamefont{Vist}} \bibnamefont{and}
  \bibinfo{author}{\bibfnamefont{J.~H.} \bibnamefont{Davis}},
  \bibinfo{journal}{Biochemistry} \textbf{\bibinfo{volume}{29}},
  \bibinfo{pages}{451} (\bibinfo{year}{1990}{\natexlab{b}}).

\bibitem[{\citenamefont{Essman et~al.}(1995)\citenamefont{Essman, Perera,
  Berkowitz, Darden, and Pedersen}}]{essman.perera.ea:smooth}
\bibinfo{author}{\bibfnamefont{U.}~\bibnamefont{Essman}},
  \bibinfo{author}{\bibfnamefont{L.}~\bibnamefont{Perera}},
  \bibinfo{author}{\bibfnamefont{M.~L.} \bibnamefont{Berkowitz}},
  \bibinfo{author}{\bibfnamefont{H.~L.~T.} \bibnamefont{Darden}},
  \bibnamefont{and} \bibinfo{author}{\bibfnamefont{L.~G.}
  \bibnamefont{Pedersen}}, \bibinfo{journal}{J. Chem. Phys.}
  \textbf{\bibinfo{volume}{103}}, \bibinfo{pages}{8577} (\bibinfo{year}{1995}).

\bibitem[{\citenamefont{Berendsen et~al.}(1984)\citenamefont{Berendsen, Postma,
  DiNola, and Haak}}]{berendsen.postma.ea:molecular}
\bibinfo{author}{\bibfnamefont{H.~J.~C.} \bibnamefont{Berendsen}},
  \bibinfo{author}{\bibfnamefont{J.~P.~M.} \bibnamefont{Postma}},
  \bibinfo{author}{\bibfnamefont{A.}~\bibnamefont{DiNola}}, \bibnamefont{and}
  \bibinfo{author}{\bibfnamefont{J.~R.} \bibnamefont{Haak}},
  \bibinfo{journal}{J. Chem. Phys.} \textbf{\bibinfo{volume}{81}},
  \bibinfo{pages}{3684} (\bibinfo{year}{1984}).

\bibitem[{\citenamefont{Frenkel and Smit}(2002)}]{Frenkel-Smit}
\bibinfo{author}{\bibfnamefont{D.}~\bibnamefont{Frenkel}} \bibnamefont{and}
  \bibinfo{author}{\bibfnamefont{B.}~\bibnamefont{Smit}},
  \emph{\bibinfo{title}{Understanding Molecular Simulation: From Algorithms to
  Applications, 2nd edition}} (\bibinfo{publisher}{Academic Press},
  \bibinfo{address}{San Diego}, \bibinfo{year}{2002}).

\bibitem[{\citenamefont{Thijsse et~al.}(1998)\citenamefont{Thijsse, Hollanders,
  and Hendrikse}}]{thijsse.hollanders.ea:practical}
\bibinfo{author}{\bibfnamefont{B.~J.} \bibnamefont{Thijsse}},
  \bibinfo{author}{\bibfnamefont{M.~A.} \bibnamefont{Hollanders}},
  \bibnamefont{and}
  \bibinfo{author}{\bibfnamefont{J.}~\bibnamefont{Hendrikse}},
  \bibinfo{journal}{Computers In Physics} \textbf{\bibinfo{volume}{12}},
  \bibinfo{pages}{393} (\bibinfo{year}{1998}).

\bibitem[{\citenamefont{Chiu et~al.}(2002)\citenamefont{Chiu, Jakobsson, Mashl,
  and Scott}}]{chiu.jakobsson.ea:cholesterol-induced}
\bibinfo{author}{\bibfnamefont{S.~W.} \bibnamefont{Chiu}},
  \bibinfo{author}{\bibfnamefont{E.}~\bibnamefont{Jakobsson}},
  \bibinfo{author}{\bibfnamefont{R.~J.} \bibnamefont{Mashl}}, \bibnamefont{and}
  \bibinfo{author}{\bibfnamefont{H.~L.} \bibnamefont{Scott}},
  \bibinfo{journal}{Biophys. J.} \textbf{\bibinfo{volume}{83}},
  \bibinfo{pages}{1842} (\bibinfo{year}{2002}).

\bibitem[{\citenamefont{McConnell and Radhakrishnan}(2003)}]{McC03}
\bibinfo{author}{\bibfnamefont{H.~M.} \bibnamefont{McConnell}}
  \bibnamefont{and}
  \bibinfo{author}{\bibfnamefont{A.}~\bibnamefont{Radhakrishnan}},
  \bibinfo{journal}{Biochim. Biophys. Acta} \textbf{\bibinfo{volume}{1610}},
  \bibinfo{pages}{159} (\bibinfo{year}{2003}).

\bibitem[{gro(2004)}]{gromacs-manual}
\bibinfo{journal}{GROMACS manual; see www.gromacs.org}  (\bibinfo{year}{2004}).

\bibitem[{\citenamefont{de~Vries
  et~al.}(2004{\natexlab{b}})\citenamefont{de~Vries, Mark, and
  Marrink}}]{vries.mark.ea:binary}
\bibinfo{author}{\bibfnamefont{A.~H.} \bibnamefont{de~Vries}},
  \bibinfo{author}{\bibfnamefont{A.~E.} \bibnamefont{Mark}}, \bibnamefont{and}
  \bibinfo{author}{\bibfnamefont{S.~J.} \bibnamefont{Marrink}},
  \bibinfo{journal}{J. Phys. Chem. B} \textbf{\bibinfo{volume}{108}},
  \bibinfo{pages}{2454} (\bibinfo{year}{2004}{\natexlab{b}}).

\bibitem[{\citenamefont{Somerharju et~al.}(1999)\citenamefont{Somerharju,
  Virtanen, and Cheng}}]{somerharju.virtanen.ea:lateral}
\bibinfo{author}{\bibfnamefont{P.}~\bibnamefont{Somerharju}},
  \bibinfo{author}{\bibfnamefont{J.~A.} \bibnamefont{Virtanen}},
  \bibnamefont{and} \bibinfo{author}{\bibfnamefont{K.~H.} \bibnamefont{Cheng}},
  \bibinfo{journal}{Biochim. Biophys. Acta} \textbf{\bibinfo{volume}{1440}},
  \bibinfo{pages}{32} (\bibinfo{year}{1999}).

\bibitem[{\citenamefont{Bach and Wachtel}(2003)}]{bach.wachtel:phospholipid}
\bibinfo{author}{\bibfnamefont{D.}~\bibnamefont{Bach}} \bibnamefont{and}
  \bibinfo{author}{\bibfnamefont{E.}~\bibnamefont{Wachtel}},
  \bibinfo{journal}{Biochim. Biophys. Acta} \textbf{\bibinfo{volume}{1610}},
  \bibinfo{pages}{187} (\bibinfo{year}{2003}).

\bibitem[{\citenamefont{Ermak and Buckholtz}(1980)}]{ermak.buckholtz:numerical}
\bibinfo{author}{\bibfnamefont{D.~L.} \bibnamefont{Ermak}} \bibnamefont{and}
  \bibinfo{author}{\bibfnamefont{H.}~\bibnamefont{Buckholtz}},
  \bibinfo{journal}{J. Comput. Phys.} \textbf{\bibinfo{volume}{35}},
  \bibinfo{pages}{169} (\bibinfo{year}{1980}).

\bibitem[{\citenamefont{Andersen}(1980)}]{andersen:molecular}
\bibinfo{author}{\bibfnamefont{H.~C.} \bibnamefont{Andersen}},
  \bibinfo{journal}{J. Chem. Phys.} \textbf{\bibinfo{volume}{72}},
  \bibinfo{pages}{2384} (\bibinfo{year}{1980}).

\end{thebibliography}

\end{document}